\documentclass[aps,twocolumn,showkeys]{revtex4}

\usepackage{graphicx}
\usepackage{subfigure}
\usepackage{color}
\usepackage{amsmath,amssymb}
\usepackage{amsthm} 
\usepackage{epsfig}
\usepackage{xcolor}
\usepackage[colorlinks=true,allcolors=blue]{hyperref}
\usepackage{bbold}
\usepackage[T1]{fontenc}
\usepackage{young}
\usepackage{youngtab}
\usepackage{etoolbox}
\usepackage{physics}
\usepackage{braket}
\usepackage{mathtools}

\makeatletter 
\renewcommand*\env@matrix[1][\arraystretch]{%
	\edef\arraystretch{#1}%
	\hskip -\arraycolsep
	\let\@ifnextchar\new@ifnextchar
	\array{*\c@MaxMatrixCols c}}
\makeatother

\renewcommand{\arraystretch}{1.5}

\def\nn{\nonumber}
\def\ic{\mathrm{i}}

\def \bc {\begin{center}}
\def \ec {\end{center}}
\def \bi {\begin{itemize}}
\def \ei {\end{itemize}}
\def \ba {\begin{array}}
\def \ea {\end{array}}
\def \bea {\begin{eqnarray}}
\def \eea {\end{eqnarray}}
\def \be {\begin{equation}}
\def \ee {\end{equation}}

\newcommand{\ra}{\rangle}

\def\ac {\bar{\alpha}}
\def\bc {\bar{\beta}}
\def\gc {\bar{\gamma}}

\def\mb{\mathbb{m}}

\def\mbhw{\mathbb{m}_{\mathrm{hw}}}
\def\mblw{\mathbb{m}_{\mathrm{lw}}}

\begin{document}

Published in Phys. Rev. E103, 012116 (2021): \href{https://link.aps.org/doi/10.1103/PhysRevE.103.012116}{https://link.aps.org/doi/10.1103/PhysRevE.103.012116}

\title{Role of mixed permutation symmetry sectors in the thermodynamic limit of critical three-level Lipkin-Meshkov-Glick atom models}

\author{Manuel Calixto}
\email{calixto@ugr.es}
\affiliation{Department of Applied Mathematics and Institute Carlos I of Theoretical and Computational Physics, University of  Granada,
Fuentenueva s/n, 18071 Granada, Spain}
\author{Julio Guerrero}
\email{jguerrer@ujaen.es}
\affiliation{Department of Mathematics, University of Jaen, Campus Las Lagunillas s/n, 23071 Jaen, Spain}
\author{Alberto Mayorgas}
\email{albmayrey97@gmail.com}
\affiliation{Department of Applied Mathematics and Institute Carlos I of Theoretical and Computational Physics, University of  Granada,
Fuentenueva s/n, 18071 Granada, Spain}

\date{\today}

\begin{abstract}
We introduce the notion of Mixed Symmetry Quantum Phase Transition (MSQPT) 
as singularities in the transformation of the lowest-energy state  properties of a system of identical particles 
inside each permutation symmetry sector $\mu$, when some Hamiltonian control parameters $\lambda$ are varied. We use a three-level Lipkin-Meshkov-Glick (LMG) model, with 
$U(3)$ dynamical symmetry, to exemplify our construction. 
After reviewing the construction of $U(3)$ unirreps using Young tableaux and Gelfand basis, we firstly study the case of a finite number $N$ of three-level atoms, showing that some precursors (fidelity-susceptibility, level population, etc.) 
of MSQPTs appear in all permutation symmetry sectors. Using coherent (quasi-classical) states of $U(3)$ as variational states, we compute the lowest-energy density for each sector $\mu$ in the 
thermodynamic $N\to\infty$ limit. Extending the control parameter space by $\mu$, the phase diagram exhibits four distinct quantum phases in the $\lambda$-$\mu$ plane that coexist at a quadruple point. 
The ground state of the whole system belongs to the fully symmetric sector $\mu=1$ and shows a four-fold degeneracy, due to the spontaneous breakdown of the parity symmetry of the Hamiltonian. The restoration 
of this discrete symmetry leads to the formation of four-component Schr\"odinger cat states. 
\end{abstract}


\maketitle

\section{Introduction}

The role of permutation symmetry is crucial in the study of the evolution of quantum systems of identical particles (the simplest example is the classification of indistinguishable 
particles as bosons or fermions attending to their permutation properties) and should be taken into consideration, not only to simplify the problem and classify their solutions, but also at a 
deeper level. A non-trivial example, which is in the realm of many important physical models, 
is that of a number of identical particles distributed in
a set of levels and a second quantized Hamiltonian describing pair correlations. In this case, 
the tensor product Hilbert space corresponding to $N$  particles/atoms distributed among $L$ $N$-fold degenerate 
levels is $L^N$ dimensional (the number of ways to put $N$ particles in $L$ levels). Particular interesting 
cases are systems of qubits $(L=2)$ and qutrits $(L=3)$, to use the quantum information jargon. When atoms are identical, permutation 
symmetry $S_N$ allows to decompose this tensor product into a ``Clebsh-Gordan'' direct sum of  unitary irreducible representations (unirreps) of the 
unitary group $U(L)$, whose generators define the dynamical algebra of the Hamiltonian in terms of collective operators. 
Young tableaux turn out to be a useful graphical method to represent these kind of decompositions into different symmetry sectors  
and we shall make use of them in the next Sections.

When dealing with  critical and chaotic  quantum systems in the thermodynamic (classical) limit $N\to \infty$, like in quantum phase transitions (QPTs), 
only the totally symmetric sector is considered in the literature (see e.g. \cite{Kus,KusLipkin,Meredith,Casati,Saraceno}), 
which reduces the size of the original Hilbert space from $L^N$ to $N+1$ for $L=2$ (symmetric spins, qubits) or to $(N+1)(N+2)/2$ 
for $L=3$ (symmetric qutrits) and so on, that is,  the number of ways of exciting $N$ atoms with $L$ levels when order does not matter. 
This means to make the atoms/particles indistinguishable. This is a common procedure in the literature which is mostly assumed without 
a clear physical justification (usually for computational convenience). It is true that there are particular situations where restricting to the 
totally symmetric sector can be physically justified.  Namely, for the Dicke model of superradiance, 
the assumption that the $N$ two-level atoms are indistinguishable is admissible when the emitters are confined to a cavity 
volume $V\ll \ell^3$ much smaller than the scale of the wavelength $\ell$ of the optical transition. Also, in the 
analysis of the phase diagram and critical points of a QPT, the restriction to the fully symmetric sector is justified under the 
assumption that the  ground state always belongs to this sector. However, as far as we know, there is not a general proof of this fact. 
One can find arguments in the literature (see e.g. \cite{Meredith} for the study of quantum chaos in a three-level LMG  shell model) 
precluding the consideration of other permutation symmetry sectors than the totally symmetric under the argument that mixed symmetry sectors  
correspond to systems with more degrees of freedom that do not approach the classical $N\to\infty$ limit as 
``rapidly'' as the totally symmetric sector does. However, no notion of ``speed/order'' of convergence to $N\to\infty$ appears in these studies. 

In this work, we want to explore the role of mixed permutation symmetry sectors usually disregarded in the study of the thermodynamic limit 
of many-body critical quantum systems. For this purpose, we shall consider the paradigmatic and ubiquitous LMG Hamiltonian used in several fields 
(nuclear, quantum optics, condensed matter, etc.) of physics to model many-body $L$-level (usually $L=2$) systems (see e.g. \cite{octavio} and references therein). 
We shall consider $L=3$, since for $L=2$ all sectors can be reduced to the symmetric one (standard Clebsch-Gordan decomposition), although we shall give a brief 
of the $L=2$ case for pedagogical reasons. We shall classify the Hamiltonian spectrum according to different permutation symmetry sectors and we shall analyze 
the lowest-energy state inside each of this sectors, leading to the new notion of  \emph{Mixed Symmetry Quantum Phase Transition} (MSQPT) in the $N\to\infty$ limit. 
Mixed symmetry sectors correspond in general with larger phase spaces than the fully symmetric sector (except for its conjugated representation), which contains the 
ground state of the system, defining the standard QPT. This notion of MSQPT is consistent since temporal evolution does not mix different symmetry sectors. 
If the initial state lays in one of these sectors, it remains trapped there. Phase diagrams and critical points depend on the particular symmetry sector and we give 
the explicit dependence for the three-level LMG model. Firstly we make a numerical treatment for large (but finite) $N$, computing some ``precursors'' of the 
MSQPT (level populations and information-theoretic measures). Then we analyze the thermodynamic $N\to\infty$ limit by using mixed-symmetry coherent states as 
variational states for the lowest-energy state inside each sector. The variational approach provides the phase diagram for each MSQPT.

This notion of MSQPT overlaps with the existing notion of Excited State Quantum Phase Transition (ESQPT) already present in the literature \cite{ESQPT,relano}. 
ESQPT is a continuation of the concept of QPT for singularities of the ground state to singularities of the excites states and singular level densities. 
From this point of view, our lowest-energy states inside each mixed symmetry sector are in fact excited energy states of the whole system, although ESQPT generally 
make reference to excited states inside the fully symmetric representation (other mixed symmetry sectors are disregarded). Therefore, our concept of 
MSQPT differs from the ESQPT notion, although there are some formal similarities. 

The organization of the article is as follows.  In Sec.~\ref{LMGsec} the general LMG model with $L$ levels is introduced, giving its main properties and  particular 
expressions for the case $L=2$ and $L=3$. In Sec.~\ref{unirreps} the unirreps of $U(L)$ are discussed using the diagrammatic approach of Young tableaux, Weyl patterns, 
and Gelfand-Tsetlin (GT) basis to classify and label the states in each unirrep. In Sec.~\ref{finiteN} the case of a finite number $N$ of three-level atoms is 
thoroughly discussed for the LMG Hamiltonian, and the fidelity susceptibility and level population are used to detect precursors of phase transitions (that, properly speaking, take place in the 
thermodynamic limit $N\rightarrow \infty$) as the interaction/control  parameter $\lambda$ is varied. Sec.~\ref{cohsec} is devoted to the definition of coherent (quasi-classical) states for each unirrep of $U(3)$ and the computation
of expectation values of $U(3)$ generators on coherent states (the so called ``symbols''). In Sec.~\ref{phasesec}, the thermodynamic limit is performed in the expectation value of the LMG Hamiltonian on coherent states, 
thus defining a energy surface which is minimized to obtain the minimum energy inside each symmetry sector $\mu$ as a function of the control parameter $\lambda$. This defines a phase diagram in the extended 
$\lambda$-$\mu$ plane with four distinct phases that coexist at a quadruple point. We pay special attention to the totally symmetric sector $\mu=1$, where the (degenerated) ground state lives, calculating level population densities 
and studying the spontaneous breakdown of parity symmetry. After a conclusion section, Appendix \ref{app1} contains the details of the derivation of the differential realization of the 
generators of $U(3)$ and their symbols, and in Appendix \ref{app2} we explicitly calculate the exponential action of $U(3)$ Cartan  generators, that leads to parity transformations when acting on coherent states.

\section{The $L$-level LMG model}\label{LMGsec}

Many models describing pairing correlations in condensed matter and nuclear physics are defined 
by a second quantized  Hamiltonian of the form
\begin{equation}
H_L=\sum_{i=1}^{L}\sum_{\mu=1}^{N}\varepsilon_i c_{i\mu}^\dag c_{i\mu}-
\sum_{i,j,k,l=1}^L\sum_{\mu,\nu=1}^N
\lambda_{ij}^{kl} c_{i\mu}^\dag c_{j\mu} c_{k\nu}^\dag c_{l\nu}\label{ham1}
\end{equation}
where $c_{i\mu}^\dag$ ($c_{i\mu}$) creates (destroys) a fermion in the $\mu$ state of a $L$-level,   
$i=1,\dots,L$, system (namely, $L$ energy levels)  with level energies $\varepsilon_i$. 
That is, the model has $N$ identical particles distributed among 
$L$ energy levels, each of which is $N$-fold degenerate. The two-body  residual interactions  (with strength $\lambda$) scatter
pairs of particles between the $L$ levels without changing the
total number of particles. For hermitian $H_L$ we have $\bar{\lambda}_{ij}^{kl}=\lambda_{lk}^{ji}$. 
Defining the $U(L)$ ``quasispin'' collective operators 
\begin{equation}
S_{ij}=\sum_{\mu=1}^N c^\dag_{i\mu}c_{j\mu}
\end{equation}
the Hamiltonian \eqref{ham1} can be written as 
\begin{equation}
H_L=\sum_{i=1}^{L}\varepsilon_i S_{ii}-\sum_{i,j,k,l=1}^L
\lambda_{ij}^{kl} S_{ij}S_{kl}\label{hamUL}\:.
\end{equation}
In this article we shall adopt a $L$-level atom picture and denote by $E_{ij}=|i\rangle\langle j|$ the (Hubbard) operator describing a 
transition from the single-atom level $|j\rangle$ to the level $|i\rangle$, with $i,j=1,\dots,L$. 
The expectation values of $E_{ij}$ account for  complex polarizations or coherences for $i\not=j$ and 
occupation probability of the level $i$ for $i=j$. The  $E_{ij}$ represent the $L^2$ generators (step operators) 
of $U(L)$ (or $GL(L;\mathbb{C})$ to be more precise) in the fundamental $L\times L$ representation, whose (Cartan-Weyl) matrices are 
$\langle l|E_{{ij}}|k\rangle=\delta _{{il}}\delta _{{jk}}$ (entry 1 in row $i$, 
column $j$ and zero elsewhere)  with commutation relations 
\be
\left[E_{{ij}},E_{{kl}}\right]=\delta _{{jk}} E_{{il}} -\delta _{{il}} E_{{kj}}.\label{commurel}
\ee
Denoting by $E_{ij}^\mu$, $\mu=1,\dots,N$ the embedding of the single $\mu$-th atom  $E_{ij}$ operator into the $N$-atom Hilbert space (namely, 
$E_{ij}^3=\mathbb{1}_L\otimes \mathbb{1}_L\otimes E_{ij}\otimes \mathbb{1}_L$ for $N=4$, with $\mathbb{1}_L$ the $L\times L$ identity), the collective quasispin 
operators are 
\be 
S_{ij}=\sum_{\mu=1}^N E_{ij}^\mu.\label{collectiveS}
\ee
They constitute the $U(L)$ dynamical algebra of our system, with the same commutation relations 
as those of $E_{ij}$ in \eqref{commurel}. 

We shall eventually particularize to $L=3$-level atoms (qutrits) for concrete calculations. The best known case is the original $L=2$ levels LMG schematic shell model \cite{lipkin1,lipkin2,ring} 
to describe the quantum phase transition from spherical to deformed shapes in nuclei. This model 
assumes that the nucleus is a system of fermions which can occupy two levels $i=1,2$ with the same
degeneracy $N$, separated by an energy $\varepsilon=2\varepsilon_2=-2\varepsilon_1$. It can also describe
a system of $N$ interacting two-level identical atoms (``qubits''), or an anisotropic XY
Ising model 
\begin{equation}
H_{XY}=  \varepsilon \sum_{\mu=1}^N \sigma_\mu^z 
+  \sum_{\mu<\nu} \lambda_x \sigma_\mu^x  \sigma_\nu^x  +  \sum_{\mu<\nu} \lambda_y\sigma_\mu^y  \sigma_\nu^y\:,
\label{Hlmgeneralpauli}
\end{equation}
in an external transverse magnetic field $\varepsilon$ with infinite-range constant
interactions. In terms of the $U(2)$ angular momentum $\vec{J}$ collective operators 
$J_+=S_{21}, J_-=S_{12}$ and $J_z=\frac{1}{2}(S_{22}-S_{11})$, the two-level LMG schematic shell model Hamiltonian reads \cite{lipkin1,lipkin2}:
\begin{equation}
H_2 = \varepsilon J_z+\frac{\lambda_1}{2}(J_+^2+J_-^2)+\frac{\lambda_2}{2}(J_+J_-+J_-J_+)\:.
\label{hamU2}
\end{equation}
The $\lambda_1$ interaction term annihilates pairs of particles in one level and creates pairs in the other level, and the $\lambda_2$
term scatters one particle up while another is scattered down. The total number of particles $N$  and the squared angular 
momentum $\vec{J}^2=j(j+1)$  are conserved. Since the Hamiltonian is symmetric under 
permutation of particle labels, it does not couple different angular momentum sectors $j=N/2,N/2-1,\dots, 1/2$ or $0$ (for odd or even $N$, respectively), 
with dimensions $2j+1$. Therefore, permutation 
symmetry reduces the size of the largest matrix to be diagonalized from $2^N$ to $N+1$. As already said, quantum calculations are usually restricted to this 
$(N+1)$-dimensional  totally symmetric subspace under the assumption that the $N$ two-level particles are indistinguishable. 
Therefore, the Hilbert space is spanned by Dicke states $|j,m\rangle, m=-j,\dots,j$, where the eigenvalue 
$m$ of $J_z$ gives the number $n=m+j$ of excited particle-hole pairs or atoms. The Hamiltonian $H_2$
also commutes with the parity operator ${\Pi}=e^{i\pi(J_z+j)}$, so that temporal evolution 
does not connect states with different parity. This parity symmetry $\mathbb{Z}_2$ is spontaneously broken in the thermodynamic $N\to\infty$ limit, giving rise to a 
degenerate ground state.

In this article we shall tackle the less familiar case of $L=3$ level LMG model, of which there are some studies in the literature (see e.g. 
\cite{Kus,KusLipkin,Meredith,Casati,Saraceno}). As in Ref. \cite{Meredith}, we shall choose for simplicity vanishing interactions for particles 
in the same level and equal interactions for particles in different levels [similar to setting  $\lambda_2=0$ in \eqref{hamU2}]. More 
explicitly, we take 
\be\lambda_{ij}^{kl}=\frac{\lambda}{N(N-1)}\delta_{ik}\delta_{jl}(1-\delta_{ij})\ee
in \eqref{hamUL}, where we are 
dividing two-body interactions by the number of particle pairs $N(N-1)$ to make the Hamiltonian an intensive quantity (energy density) 
since we are interested in the thermodynamical limit $N\to\infty$. 
We shall also place the levels symmetrically about the level $i=2$ and write the 
intensive energy splitting per particle $\varepsilon_3=-\varepsilon_1=\epsilon/N$ and $\varepsilon_2=0$. 
Therefore, our  Hamiltonian density will be:
\begin{equation}
H=H_3=\frac{\epsilon}{N}(S_{33}-S_{11})-\frac{\lambda}{N(N-1)}\sum_{i\not=j=1}^3 S_{ij}^2.\label{hamU3}
\end{equation}
This Hamiltonian density is invariant under the combined interchange of levels $1\leftrightarrow 3$ and $\epsilon\to -\epsilon$. 
We shall take $\epsilon>0$,  for simplicity, measure energy in $\epsilon$ units, and discuss the 
energy spectrum and the phase diagram in terms of the control parameter $\lambda$. 
The existence of an interesting parity symmetry (like in the two-level case) also deserves attention. Indeed, 
this symmetry of the Hamiltonian has to do with the fact that the interaction only scatters pairs of particles, thus conserving 
the parity $\Pi_i=\exp(\ic \pi S_{ii})$, even (+) or odd ($-$),  of the population $S_{ii}$ in each level $i=1,\dots,L$. For concreteness, we shall restrict to the case $L=3$  (see also  \cite{Meredith}). 
Therefore, there are four different Hamiltonian  
matrices  identified by $(\Pi_1,\Pi_2,\Pi_3)$
\begin{equation}
(+,+,+), (+,-,-), (-,-,+), (-,+,-) \nonumber
\end{equation}
for even $N$ and 
\begin{equation}
(-,-,-), (-,+,+), (+,+,-), (+,-,+) \nonumber
\end{equation}
for odd $N$. This discrete symmetry corresponds to the finite group $\mathbb{Z}_2\times\mathbb{Z}_2\times\mathbb{Z}_2$ with the constraint 
$\Pi_1\Pi_2\Pi_3=e^{i\pi N}$. It is spontaneously broken in the thermodynamic limit and gives rise to a highly degenerated ground state, 
as compared to the $L=2$ case (see Section \ref{phasesec} and Appendix \ref{app2} for more 
details).

As for the two-level case of $H_2$, with regard the rotation group $U(2)$ and Dicke states, the Hamiltonian matrix of $H_3$ is block diagonal when 
the basis vectors are adapted to irreducible representations of the Lie group $U(3)$. 
Let us make a brief summary  the general decomposition of the Hilbert space of $N$ $L$-level atoms into $U(L)$ irreducibles. We shall restrict ourselves to $L=2$ 
(qubits) and $L=3$ (qutrits) for the sake of simplicity, although the procedure can be easily extrapolated to general $L$. Those readers acquainted with 
this language can skip to the next section.

\section{U(L) unirreps: Young tableaux, Gelfand basis and matrix elements}
\label{unirreps}

Le us symbolize the fundamental $L\times L$ representation of $U(L)$ by the Young box $\yng(1)$ . The single atom states are represented by 
Weyl patterns/tableaux by filling in the boxes with integers $i=1,\dots,L$ (the number of levels). For example, for $L=2$
\[\Yvcentermath1 \young(1)=|1\rangle, \quad\young(2)=|2\rangle,\]
symbolize the (spin) doublet, and for $L=3$
\[\Yvcentermath1 \young(1)=|1\rangle,\quad \young(2)=|2\rangle, \quad\young(3)=|3 \rangle,\]
symbolize the triplet. The unitary group $U(L)$ is represented in this space by the fundamental ($L$-dimensional) representation. For $L=2$, 
unitary matrices  $V\in U(2)$ can be obtained by applying Gram-Schmidt orthonormalization procedure to the 
columns of the triangular matrix $T$ as
\be
T=\left(
\begin{array}{cc}
 1 & 0  \\
 \alpha  & 1 
\end{array}
\right)\xrightarrow{\text{G-S}} V=\left(\begin{array}{ccc}
 \frac{1}{\sqrt{\ell}} & \frac{-\bar{\alpha}}{\sqrt{\ell}}   \\
 \frac{\alpha}{\sqrt{\ell}}   &  \frac{1}{\sqrt{\ell}}
\end{array}\right)\label{GSU2}
\ee
where  $\alpha$ is a complex number and $\ell=|T^\dag T|_1=1+\alpha  \bar{\alpha }$ is the  leading principal minor  of 
$T^\dag T$. The addition of two phases (complex numbers $u_1$ and $u_2$ of modulus 1), as $U=V\cdot \mathrm{diag}(u_1,u_2)$, 
completes the parametrization of $U(2)$ by the coordinates: $\alpha, u_1$ and $u_2$. For $L=3$ levels, 
unitary matrices $U=V\cdot\mathrm{diag}(u_1,u_2,u_3)\in U(3)$ can be  constructed following a similar procedure, with
\be
T=\left(
\begin{array}{ccc}
 1 & 0 & 0 \\
 \alpha  & 1 & 0 \\
 \beta  & \gamma  & 1 \\
\end{array}
\right)\xrightarrow{\text{G-S}} V=\left(\begin{array}{ccc}
 \frac{1}{\sqrt{\ell_1}} & \frac{-\bar{\alpha }-\gamma  \bar{\beta }}{\sqrt{\ell_1\ell_2}} &   \frac{-\bar{\beta }+\bar{\alpha } \bar{\gamma }}{\sqrt{\ell_2}}   \\
 \frac{\alpha}{\sqrt{\ell_1}}   & \frac{1+\beta  \bar{\beta }-\alpha  \gamma  \bar{\beta }}{\sqrt{\ell_1\ell_2}}  & \frac{-\bar{\gamma }}{\sqrt{\ell_2}}  \\
 \frac{\beta}{\sqrt{\ell_1}}   & \frac{\gamma-\beta  \bar{\alpha }+\gamma\alpha    \bar{\alpha }  }{\sqrt{\ell_1\ell_2}}   &  \frac{1}{\sqrt{\ell_2}}
\end{array}\right)\label{GSU3}
\ee
where $\alpha,\beta,\gamma$ are complex numbers and 
\bea \ell_1&=&|T^\dag T|_1=1+\alpha  \bar{\alpha }+\beta  \bar{\beta },\label{lengths}\\
\ell_2&=&|T^\dag T|_2=1+\gamma  \bar{\gamma }+(\beta -\alpha  \gamma ) \left(\bar{\beta }-\bar{\alpha } \bar{\gamma }\right),\nonumber
\eea
are the leading principal minors of order 1 and 2  of $T^\dag T$ (or the squared inverse leading principal minors of order 
1 and 2  of $V$), which play an important role in computing coherent state expectation values (see later on Section \ref{cohsec}).

For $N$ identical $L$-level atoms, the  $L^N$-dimensional Hilbert space is the $N$-fold tensor product
\[\begin{gathered}\yng(1)\end{gathered}\:\otimes\:\stackrel{N\ \mathrm{times}}{\dots}\:\otimes\: \begin{gathered}\yng(1)\end{gathered}\,.\]
The tensor product representation of $U(L)$ is now reducible and the invariant subspaces are graphically represented by Young frames of 
$N$ boxes
 \be \overbrace{\young(\quad \cdots\cdots\cdots\cdots\cdots\quad,\quad\vdots\vdots\vdots\quad,\quad\cdots\quad)}^{h_1}\label{tensorprod}\ee
of shape  $h=[h_1,\dots,h_L]$, with  $h_1\geq \dots \geq h_L$, $h_i$ the number of boxes in row $i=1,\dots,L$ and $h_1+\dots +h_L=N$. 
For example, let us consider the case of 
$N=3$ two-level identical atoms (three qubits in the quantum information jargon). The Hilbert space is the $2^3=8$-dimensional   
3-fold tensor product of the 2-dimensional Hilbert 
space of a single atom. In Young diagram notation, the Clebsch-Gordan direct sum decomposition of this 3-fold tensor product gives (dimensions are displayed 
on the top)
\[ \Yvcentermath1 \stackrel{2}{\yng(1)}\otimes \stackrel{2}{\yng(1)}\otimes \stackrel{2}{\yng(1)}\: = 
\:\stackrel{4}{\yng(3)}\oplus \:2\stackrel{2}{\yng(2,1)}\]
which is the analogous of the usual coupling of three spin-1/2 yielding a spin 3/2 and two spins 1/2. Note that the representations $\Yvcentermath1\yng(2,1)$ and 
$\yng(1)$ are equivalent from the point of view of $SU(2)$. This procedure can be iterated combining $N$ doublets (spin-1/2) 
to obtain the Clebsch-Gordan decomposition series (Catalan's triangle)
\be
2^{\otimes N}=\bigoplus_{k=0}^{\left \lfloor{N/2}\right \rfloor}M_k(N+1-2k),\quad M_k=\frac{N+1-2k}{N+1}\tbinom{N+1}{k}, \label{Catalan}
\ee
where $\left \lfloor{N/2}\right \rfloor$ is the integer floor function. That is, the angular momentum $j_k=\frac{N}{2}-k$ appears 
with multiplicity $M_k$. Note that the fully symmetric representation $k=0$ always appears with multiplicity $M_0=1$.
For the case of  $N=4$ qutrits, the direct-sum decomposition of the $3^4=81$-dimensional  4-fold tensor product into $U(3)$ irreducibles 
gives
\bea &\Yvcentermath1 \stackrel{3}{\yng(1)}\otimes \stackrel{3}{\yng(1)}\otimes \stackrel{3}{\yng(1)} \otimes \stackrel{3}{\yng(1)}
\:=\: & \label{CGdecompN4}\\ &\Yvcentermath1 \stackrel{15}{\yng(4)}\oplus\: {3}\stackrel{15}{\yng(3,1)}\oplus\: {2}\stackrel{6}{\yng(2,2)}   
\oplus\: 3\stackrel{3}{\yng(2,1,1)}\:.&\nonumber
\eea
The last Young frame in the previous decomposition  is equivalent to $\yng(1)$ from the point of view of $SU(3)$. In general, $h=[h_1,h_2,h_3]$ is 
equivalent to $h'=[h_1-h_3,h_2-h_3,0]$ from the point of view of $SU(3)$.

A Weyl pattern (a Young frame filled up with level labels $i=1,\dots,L$) is said to be in the semistandard form  (or column strict) 
if the sequence of labels is non-decreasing from the left to the right, 
and strictly increasing from the top to the bottom. For example, for a Young frame of shape $h=[3,2,1]$ ($N=6$ atoms), 
the following Weyl pattern
\be
 \young(112,23,3)\label{ytsf}
\ee
is in the semistandard form. The dimension of the representation $h$ coincides with the number of Weyl patterns in the semistandard form that 
one can construct. The weight or content of a Weyl pattern is a vector $w=(w_1,\dots,w_L)$ whose components $w_k$ are the population of level $k$ 
[the eigenvalues of the quasispin operators $S_{kk}$ in \eqref{collectiveS}], with $w_1+\dots+w_L=N$. 
For example, the weight of \eqref{ytsf} is $w=(2,2,2)$.

Gelfand-Tsetlin (GT) patterns   (see e.g \cite{Barut}) are a convenient way to label Weyl patterns in semistandard form (i.e., quantum states of an irreducible representation of $U(L)$ with label $h$). For example, for $L=2$, each irreducible subspace of $U(2)$ is spanned by the GT basis vectors
\be
|\mathbb{m}\rangle=\left|\begin{matrix} m_{12}=h_1 && m_{22}=h_2\\ & m_{11} &\end{matrix}\right>, \quad h_2\leq m_{11}\leq h_1.  \label{GelfandU2}
\ee
The equivalence with more standard $SU(2)$ angular momentum $j$ or Dicke states  $\{|j,m\rangle, -j\leq m\leq j \}$ is 
\be
|j=\frac{h_1-h_2}{2},m=m_{11}-\frac{h_1+h_2}{2}\rangle,\label{DickeGT}
\ee
with $h_1+h_2=N$ the linear Casimir eigenvalue of $U(2)$.  Note that two $U(2)$ irreps, $h=[h_1,h_2]$ and  $h'=[h'_1,h'_2]$, with the same angular momentum $j=\frac{h_1-h_2}{2}=\frac{h'_1-h'_2}{2}$ 
are equivalent under the point of view of $SU(2)$. In particular $h=[h_1,h_2]$ is equivalent to $h'=[h_1-h_2,0]$ under SU(2) (same angular momentum). The totally symmetric irrep $h=[N,0]$ (depicted 
by a Young frame with a single row of $N$ boxes) has the higher angular momentum $j=N/2$ in the Clebsch-Gordan sum decomposition of the $N$-fold tensor product \eqref{Catalan}. 
For $L=3$-level atoms, unirreps of $U(3)$ 
of shape/label $h=[h_1,h_2,h_3]$ are spanned by GT basis vectors 
\be |\mb\ra=\left|\begin{matrix}[1] m_{13}=h_1 & & m_{23}=h_2 & &  m_{33}=h_3 \\ 
 &  m_{12} && m_{22}  &  \\ & &  m_{11} &&   
\end{matrix}\right\rangle,\quad \label{GTU3}
\ee
which are subject to the betweenness conditions:
\bea
 &h_1\geq m_{12}\geq h_2,\quad h_2\geq m_{22}\geq h_3, & \nonumber \\ & m_{12}\geq m_{11} \geq m_{22}. &    \label{betweenU3}
 \eea
The dimension of the carrier Hilbert space $\mathcal H_h$ of an irrep of $U(3)$ of shape $h$ is then 
\bea
\mathrm{dim}(\mathcal H_h)&=& \sum _{{m_{12}}={h_2}}^{h_1} \sum _{{m_{22}}={h_3}}^{h_2} \sum _{{m_{11}}={m_{22}}}^{{m_{12}}} 1 \label{dimension}\\ &=&\frac{1}{2}(1+h_1-h_2)(2+h_1-h_3)(1+h_2-h_3).
\nonumber
\eea

The connection between Weyl and GT patterns is the following. Denoting  by $n_{k i}$ the number of times that the level $i$ appears in the row $k$ (counting downwards) of a Weyl pattern
(that is, the population of level $i$ in the row $k$), the corresponding GT labels are  $m_{k j}=\sum_{i=1}^{j}n_{k i}$. If we denote the GT pattern \eqref{GTU3} by its rows: $\mb=\{m_3,m_2,m_1\}$ with $m_3=[m_{13}, m_{23}, m_{33}]$, $m_2=[m_{12},m_{22}]$ and $m_1=[m_{11}]$, then $m_{3}$ is  
directly read off the shape of the Weyl pattern,  $m_2$    is read off the shape 
that remains after all boxes containing label 3 are removed and, finally,  $m_{1}$    is read off the shape that remains after all remaining boxes containing label 2 are removed. In the example \eqref{ytsf} we have the correspondence
\be
 \Yvcentermath1  \young(112,23,3)= \left|\begin{matrix}[1] 3 & & 2 & &  1 \\ 
 &  3 && 1  &  \\ & &  2 &&   \\
\end{matrix}\right\rangle.\label{gytsf}
\ee
The population of level $k$ (the weight component $w_{k}$) is then directly computed from a GT pattern $|\mb\ra$ as  $w_k=\bar{m}_k-\bar{m}_{k-1}$, where we denote by 
 $\bar{m}_k=\sum_{i=1}^k m_{ik}$, the average of row $k$ of the pattern $\mb$ (one sets $\bar{m}_0\equiv 0$ by convention). Therefore, the  action of diagonal operators $S_{kk}$ on 
an arbitrary GT vector $|\mb\ra$ is
\be
S_{kk}|\mb\ra=w_{k}|\mb\ra=(\bar{m}_k-\bar{m}_{k-1})|\mb\ra. \label{coefdiag}
\ee
For example, the weight of the GT vector \eqref{GelfandU2} is $(w_1,w_2)=(m_{11},N-m_{11})$ or $(\frac{N}{2}+m,\frac{N}{2}-m)$ in terms of the angular momentum third component $m$ in \eqref{DickeGT}.

A state $|\mb'\ra$ is said to be of lower weight $w'$ than $|\mb\ra$ if the 
first non-vanishing coefficient of $w-w'$ is positive.  This is called the lexicographical rule. In more physical but less precise  terms, populating lower levels $k$ increases 
the weight $w$. It is clear that  the highest weight (HW) vector $|\mb_\mathrm{hw}\rangle$ has weight $w=(h_1,h_2,h_3)$. In GT notation, the HW vector of an irrep $h$ of $U(3)$ corresponds to
\be
 |\mbhw\ra=\left|\begin{matrix}[1] h_1 & & h_2 & &  h_3 \\ 
 &  h_1 && h_2  &  \\ & &  h_1 &&   \\
\end{matrix}\right\rangle= \Yvcentermath1  \young(1\cdots\cdots\cdots\cdots\cdots  1,2\cdots\cdots\cdots 2,3\cdots 3)
\ee
Analogously, the lowest weight (LW) vector has weight $w=(h_3,h_2,h_1)$ and is given by
\be
 |\mblw\ra=\left|\begin{matrix}[1] h_1 & & h_2 & &  h_3 \\ 
 &  h_2 && h_3  &  \\ & &  h_3 &&   \\
\end{matrix}\right\rangle= \Yvcentermath1  \young(1\cdots12\cdots23\cdots3,2\cdots23\cdots3,3\cdots3)
\ee

In general, all states of the  representation $h=[h_1,\dots,h_L]$ of  $U(L)$  can be obtained from a HW vector $|\mb_\mathrm{hw}\rangle$ by applying lowering operators $S_{jk}, j>k$, or from a LW vector $|\mb_\mathrm{lw}\rangle$ by applying rising operators $S_{jk}, j<k$. Indeed, 
let us denote by $\mathbb{e}_{jk}$ the pattern with 1 at place $(j,k)$ and zeros elsewhere. The action of step 1 lowering $S_{-k}\equiv S_{k,k-1}$ and rising operators 
$S_{+k}\equiv S_{k-1,k}$ is given by (see e.g. \cite{Barut}) 
\be
S_{\pm k}|\mb\ra=\sum_{j=1}^{k-1} c^\pm_{j,k-1}(\mb)|\mb\pm\mathbb{e}_{j,k-1}\ra,\label{coefact}
\ee
with coefficients
\begin{widetext}
\be
c_{j,k-1}^\pm(\mb)=\left(-\frac{\prod_{i=1}^N(m'_{ik}-m'_{j,k-1}+\frac{1\mp 1}{2})\prod_{i=1}^{k-2}(m'_{i,k-2}-m'_{j,k-1}-\frac{1\pm 1}{2})}{\prod_{i\not=j}(m'_{i,k-1}-m'_{j,k-1})(m'_{i,k-1}-m'_{j,k-1}\mp 1)}\right)^{1/2},\label{coef}
\ee
\end{widetext}
where $m'_{ik}=m_{ik}-i$ and $c_{j,k-1}^\pm(\mb)\equiv 0$ whenever any indeterminacy arises. In fact, from the commutation relations 
\bea
[S_{ii},S_{jk}]&=&\delta_{ij}S_{ik}-\delta_{ik}S_{ji} \nonumber\\ \Rightarrow S_{ii} S_{jk}|\mbhw \rangle&=&(m_{iN}+\delta_{ij}-\delta_{ik})S_{jk}|\mbhw \rangle,\label{commuw}
\eea
the weight $w'$ of $S_{-k}|\mb\rangle$ is given by
\be
S_{ii}S_{-k}|\mb\rangle=(w_i+\delta_{i,k}-\delta_{i,k-1})S_{-k}|\mb\rangle=w'_iS_{-k}|\mb\rangle,
\ee
and therefore, $S_{-k}|\mb\rangle$ becomes of lower weight than $|\mb\ra$ since the first non-vanishing coefficient of $w-w'$ is \\$(w-w')_{k-1}=1>0$. 
Applying recursion formulas 
\begin{align}
	S_{i,i-l}&=[S_{i,i-1},S_{i-1,i-l}],\nonumber \\  
	S_{i-l,i}&=[S_{i-l,i-1},S_{i-1,i}]\quad l>0,\label{recurrence}
\end{align}
one can obtain any non diagonal operator $S_{ij}$ matrix element from \eqref{coefact}. In particular, the HW vector verifies 
\be S_{ij}|\mbhw \rangle=h_i\delta_{ij} |\mbhw \rangle\quad\forall i\leq j, 
\ee
whereas the action of lowering operators  $S_{ij}, i>j$, is given by (\ref{coefact},\ref{coef},\ref{recurrence}). From the definition \eqref{coef} one can prove that 
\be
c^\pm_{j,k-1}(\mb)=c^{\mp}_{j,k-1}(\mb\pm\mathbb{e}_{j,k-1}),
\ee
which means that $S_{+k}^\dag=S_{-k}$. Also, applying induction and the recurrence formulas \eqref{recurrence}, one obtains $S_{k,k-h}^\dag=S_{k-h,k}$. 
As a particular case, using the equivalence \eqref{DickeGT} between GT and Dicke vectors for $U(2)$, one can recover the usual  angular momentum, $J_z=(S_{22}-S_{11})/2$, $J_+=S_{21}$ and $J_-=S_{12}$,   matrix elements 
\bea
\langle j,m'|{J}_z|j,m\rangle&=&m\delta_{m',m},\label{so3j}\\
\langle j,m'|{J}_\pm|j,m\rangle&=&\sqrt{(j\mp m)(j\pm m+1)}\delta_{m',m\pm 1}. \nonumber
\eea
from the general expressions  (\ref{coefdiag},\ref{coefact},\ref{coef}). 

With all this whole mathematical arsenal, we are now ready to tackle  the analysis of the Hamiltonian \eqref{hamU3} spectrum according to permutation symmetry, and the structure of the low-energy states inside each symmetry sector. 
 
\section{Symmetry classification of Hamiltonian eigenstates for a finite number of 3-level atoms and QPT precursors}
\label{finiteN}

 Let us firstly analyze the spectrum of the noninteracting  free Hamiltonian part $H^{(0)}=\frac{\epsilon}{N}(S_{33}-S_{11})$ of the LMG Hamiltonian \eqref{hamU3}. 
 For level splitting $\epsilon>0$, the lowest-energy (ground) state coincides with  the highest-weight state 
 \begin{equation}
|\psi_0\rangle= |\mbhw\ra=\Ket{\begin{matrix}[1] N& & 0 & &  0 \\ 
 &  N && 0  &  \\ & &  N &&   
\end{matrix}}= \Yvcentermath1  \overbrace{\young(1\cdots  1)}^N\label{hwv}
\end{equation}
of the fully symmetric representation  $h=[N,0,0]$. That is, all $N$ atoms are placed at the level $i=1$. The energy density is then  $E_0=-\epsilon$.  
The excited states correspond to energy densities $E_n=(n-N)\epsilon/N, n=1,\dots,2N$. The highest-energy $E_{2N}$ state corresponds to the lowest-weight vector
\be
 |\psi_{2N}\ra=|\mblw\ra=\left|\begin{matrix}[1] N & & 0 & &  0 \\ 
 &  0 && 0  &  \\ & &  0 &&   \\
\end{matrix}\right\rangle= \Yvcentermath1  \overbrace{\young(3\cdots 3)}^N\label{lwv}
\ee
of the fully symmetric representation  $h=[N,0,0]$. That is, all $N$ atoms are placed at the level $i=3$. These free Hamiltonian eigenvalues $E_n$ 
are highly degenerated, except for $E_0$ and $E_{2N}$. For example, the states 
\[ \young(1\cdots12) \quad \mathrm{and} \quad  \Yvcentermath1   \young(1\cdots1,2)\:,\] 
which belong to different symmetry sectors, have the same energy $E_1$ (first excited energy level). The eigenvector composition and degeneracy of higher excited states is a 
bit more involved. Note that all GT vectors $|\mb\ra$ in \eqref{GTU3} are eigenvectors of the free Hamiltonian density  $H^{(0)}$ and their eigenvalues can be easily calculated as
\be
E_\mb=\frac{\epsilon}{N}(w_3-w_1)=\frac{\epsilon}{N}(N-m_{11}-m_{12}-m_{22}).
\ee
Therefore, the degeneracy of each energy level coincides with the total number of GT patterns $\mb$ with a common value of $m_{11}+m_{12}+m_{22}$. The highest- and the lowest-energy levels 
correspond to $m_{11}+m_{12}+m_{22}=0$ and $m_{11}+m_{12}+m_{22}=2N$, respectively, and they have degeneracy 1, coinciding with the lowest- and the highest-weight vectors in \eqref{lwv} and \eqref{hwv}, 
respectively.

\begin{figure}[h]
\begin{center}
\includegraphics[width=\columnwidth]{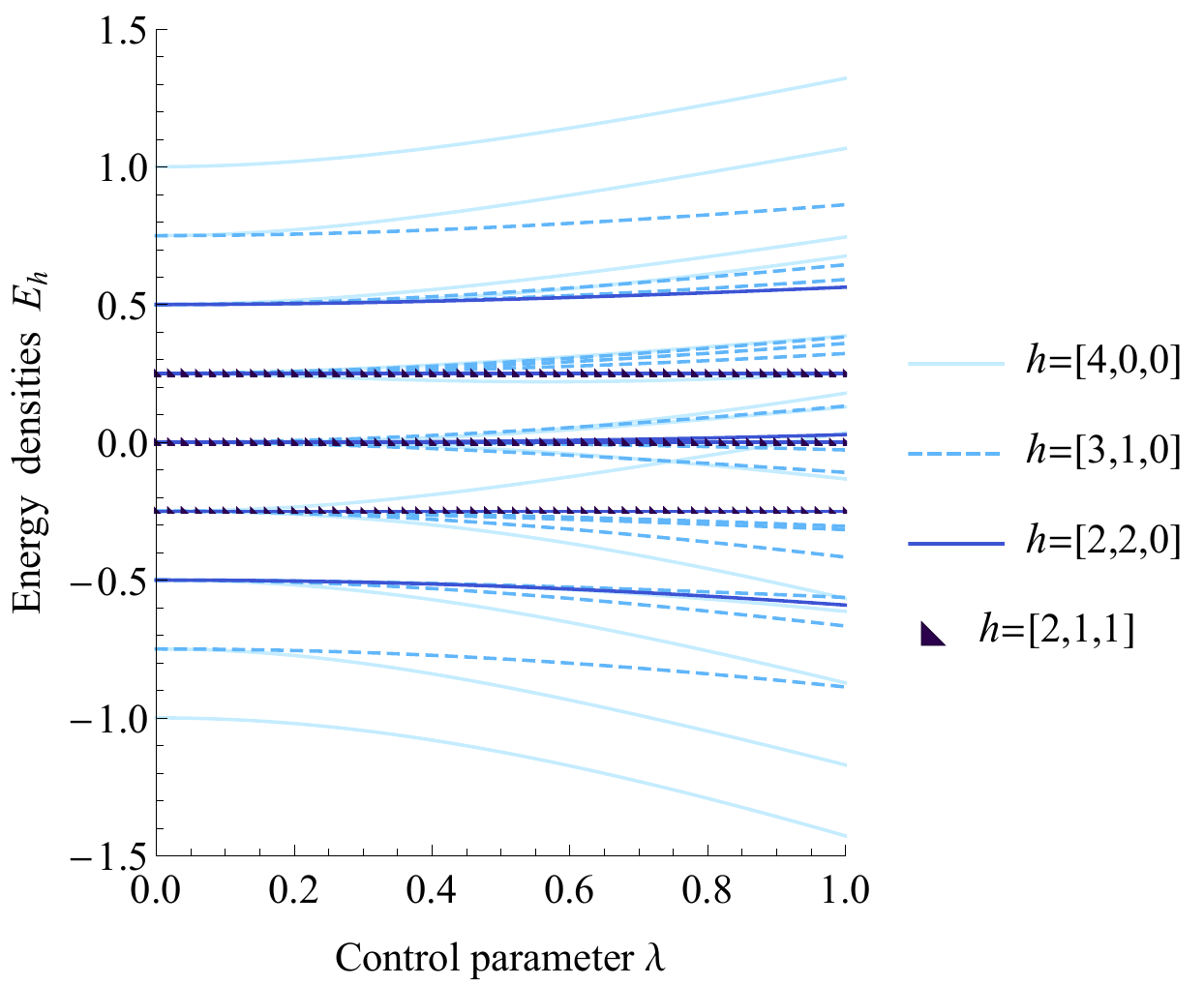}
\end{center}
\caption{LMG Hamiltonian energy density spectrum, for $N=4$ identical $L=3$ level atoms, as a function of the interacting control parameter $\lambda$ (both in $\epsilon$ units). Energy 
curves associated to the four different symmetry sectors $h$, depicted in \eqref{CGdecompN4}, are plotted with different color. The free Hamiltonian ($\lambda=0$) eigenvalues  are 
highly degenerated, the corresponding eigenspaces containing vectors belonging to different symmetry sectors $h$. This degeneracy is partially lifted when the two-body 
interaction  ($\lambda\not=0$) is introduced, giving rise to the appearance of energy bands.}
\label{fig0}
\end{figure}

This degeneracy is partially lifted when the two-body interaction [with coupling constant $\lambda$, like in \eqref{hamU3}] is introduced. This interaction affects each permutation symmetry sector in a different manner,  so that energy bands emerge in the interacting Hamiltonian 
spectrum, as we can perceive in Figure \ref{fig0} for $N=4$ identical 3-level atoms. Excited states belong not only to the fully symmetric representation but all symmetry sectors are involved. Therefore, 
if we are in a physical situation where our identical atoms are not necessarily indistinguishable, we should not disregard symmetry sectors other than the fully symmetric, since they play an important role in the analysis of excited states. 

\begin{figure}[h]
\begin{center}
\includegraphics[width=\columnwidth]{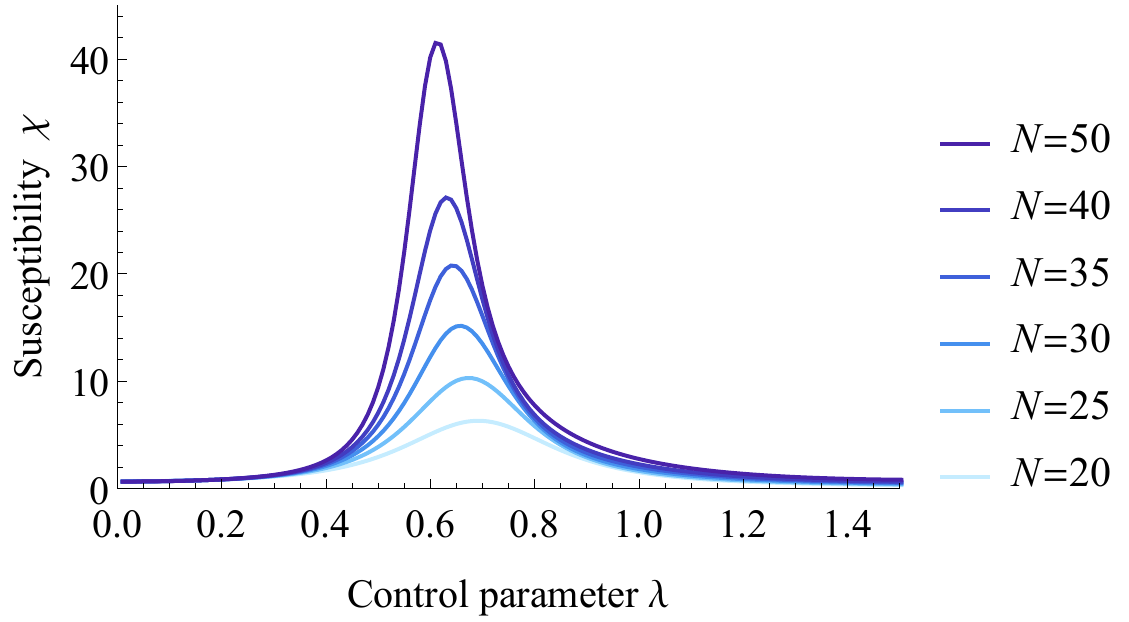}
\end{center}
\caption{Susceptibility $\chi_\psi$ of the ground (fully symmetric) state $\psi$ of the LMG Hamiltonian \eqref{hamU3} as a function of $\lambda$ for increasing values of  the number of atoms $N$. 
A step  $\delta\lambda= 0.01$ has been used. The analysis divines the existence of a QPT at a  critical point around $\lambda^{(0)}\simeq 0.6$. We use $\epsilon$ units for $\lambda$.}
\label{fig1}
\end{figure}

Since Hamiltonian evolution does not mix different symmetry sectors $h$, we are interested in the analysis of critical phenomena occurring inside each Hilbert subspace $\mathcal{H}_h$ 
corresponding to the carrier space of an irrep $h$ of  $U(3)$. Therefore, we shall select the lowest-energy vector $|\psi_0^h\rangle$ inside each $\mathcal{H}_h$ and look for 
drastic changes in its structure when varying $\lambda$ for $N\to\infty$ (thermodynamic limit). With this analysis we will introduce the concept of MSQPT in the next sections. Before, 
for finite $N$, there still are some QPT precursors which can anticipate the approximate location of critical points. The drastic change of the structure of a  state $|\psi(\lambda)\rangle$ 
in the vicinity of a critical point $\lambda^{(0)}$ can be quantified with information theoretic measures like the so called fidelity \cite{zanardi,gu,PhysRevLett.99.100603}
\[ F_\psi(\lambda,\delta\lambda)=|\langle\psi(\lambda)|\psi(\lambda+\delta\lambda)\rangle|^2,\]
which measures the overlap between two states in the vicinity ($\delta\lambda\ll 1$) of $\lambda$. The fidelity is nearly 1 far from a critical point $\lambda^{(0)}$ and drastically falls down  in the 
vicinity of $\lambda^{(0)}$, the more the higher is $N$. Instead of $F_\psi(\lambda,\delta\lambda)$, which is quite sensitive to the step $\delta\lambda$, we shall use the  so called susceptibility 
\be
\chi_\psi(\lambda,\delta\lambda)=2\frac{1-F_\psi(\lambda,\delta\lambda)}{(\delta\lambda)^2}.\label{susceptibility}
\ee
See e.g. References \cite{ma,Romera_2014,PhysRevE.92.052106} for the use of information-theoretic concepts like susceptibility and 
R\'enyi-Wehrl entropies in the 2-level LMG case and other paradigmatic QPT models. 
In Figure \ref{fig1} we represent the susceptibility of the ground (fully symmetric) state of the 3-level LMG Hamiltonian \eqref{hamU3} for increasing values of $N$. 
We see that the susceptibility is sharper and sharper as $N$ increases, divining  the existence of a QPT at a  critical point around $\lambda^{(0)}\simeq 0.6\epsilon$. 
Actually, the variational/semiclassical  $N\to\infty$ study, using coherent states \`a la Gilmore \cite{gilmore},  in sections  \ref{cohsec} and \ref{phasesec} will reveal the existence of a second order QPT at exactly 
$\lambda^{(0)}=0.5\epsilon$ (see \cite{octavio} for the variational study of the  2-level LMG case and its phase diagram).

\begin{figure}[h]
\begin{center}
\includegraphics[width=\columnwidth]{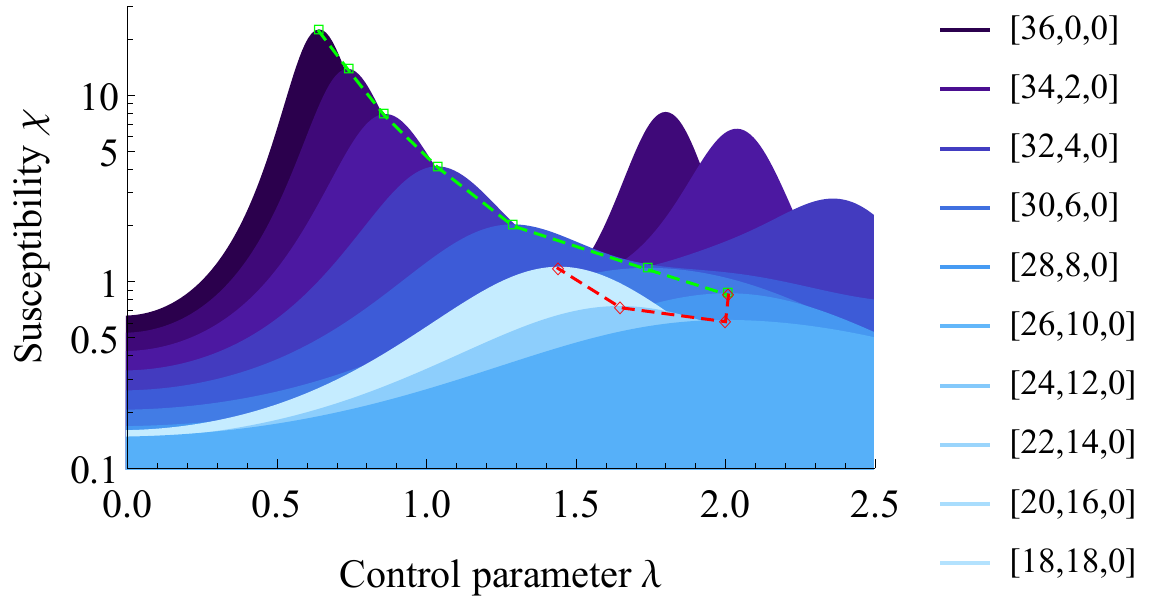}
\end{center}
\caption{Susceptibility $\chi_{\psi_0^h}$ of the  lowest-energy vector $\psi_0^h$ inside some  mixed symmetry sectors $h$ for $N=36$ atoms. Logarithmic scale. 
The dashed line interpolates between maxima of the susceptibility that are precursors of prospective critical points separating  phase I from phase II (green squares) and phase I from phase IV (red diamonds); 
see later on Figures \ref{fig4} and \ref{fig_phase}. 
The ``recoil point'' corresponds to the ``hand-gun'' unirrep $h=[24,12,0]$, where four phases will coexist (see later on Sec. \ref{phasesec}). 
We use $\epsilon$ units for $\lambda$.}
\label{fig2}
\end{figure}

The same critical phenomenon occurs for the lowest-energy state belonging to other mixed symmetry sectors $h$. In Figure \ref{fig2} we represent the susceptibility 
$\chi_{\psi_0^h}$ of the  lowest-energy vector $\psi_0^h$ inside some  mixed symmetry sectors $h$ for $N=36$ atoms. The analysis of the first maxima of the susceptibility in 
Figure \ref{fig2} indicates that the would-be critical points $\lambda^{(0)}(h)$ are shifted to the right from $h=[36,0,0]$ to $h=[24,12,0]$ and then to the left from $h=[24,12,0]$ to 
$h=[18,18,0]$. In fact, the semiclassical $N\to\infty$ analysis that we shall make in Section \ref{phasesec}, Figures \ref{fig4} and \ref{fig_phase}, indicates that the ``hand-gun'' sector 
$h=[2N/3,N/3,0]$ (we shall use this terminology for this special case, which coincides with the adjoint ``octet'' representation in quantum chromodynamics $N=3$) corresponds to a quadruple point. 
Therefore, the susceptibility is able to capture this special point. The susceptibility second maxima in Figure \ref{fig2} correspond to a new QPT that eventually takes place at $\lambda^{(0)}=1.5\epsilon$. 
We shall not discuss this last QPT until Section \ref{phasesec} since it occurs at a different scale and requires much higher values of $N$, and more computational requirements, to be 
properly captured.

\begin{figure}[h]
\begin{center}
\includegraphics[width=\columnwidth]{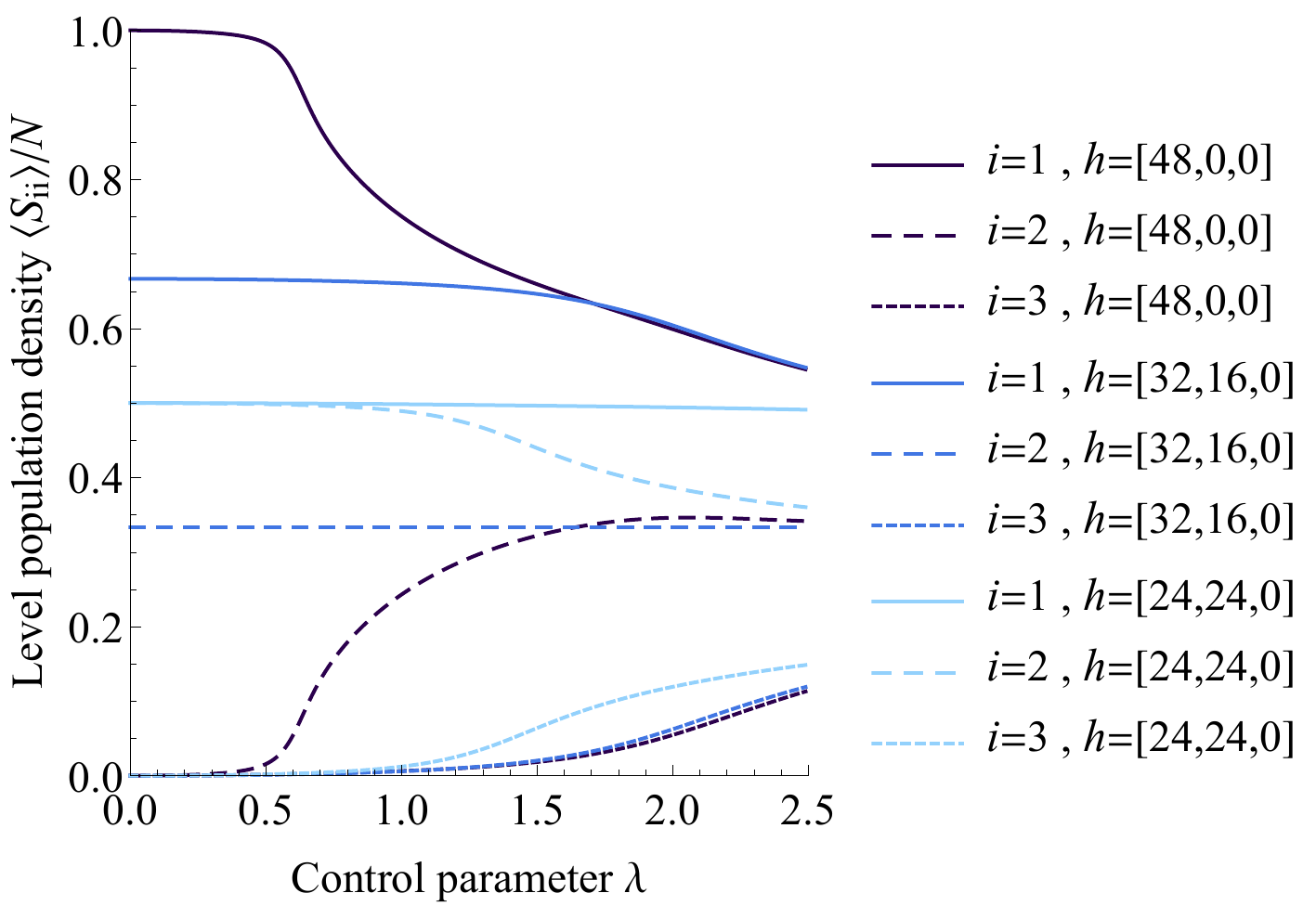}
\end{center}
\caption{Level $i=1,2,3$ population densities $\langle \psi_0^h|S_{ii}|\psi_0^h\rangle/N$ of the lowest-energy state 
$\psi_0^h$ inside each sector $h$ for $N=48$ three-level atoms and three different representative values of $h$ 
(the fully symmetric, the ``hand-gun'' and the rectangular Young tableau), as a function of $\lambda$, in $\epsilon$ units. 
Appreciable changes in the population densities can be observed at some values of $\lambda$ depending on the particular sector $h$ (see main text),  
anticipating the existence of a MSQPT in the thermodynamic limit.}
\label{fig21}
\end{figure}

Level $i=1,2,3$ population densities $\langle \psi_0^h|S_{ii}|\psi_0^h\rangle/N$, of the ground state $\psi_0^h$ inside each sector $h$, also behave as precursors of order parameters. 
In fact, Figure \ref{fig21} represents level population densities for $N=48$ three-level atoms and different symmetry sectors $h$, which include the symmetric sector $h=[48,0,0]$, 
the ``hand-gun'' sector $h=[2N/3,N/3,0]=[32,16,0]$ already commented in the previous paragraph, and the rectangular Young tableau $h=[24,24,0]$. We perceive a population change for the fully 
symmetric case around $\lambda^{(0)}=0.5\epsilon$, as already pointed out, a displacement of this critical point to the right for $h=[32,16,0]$ (the would-be quadruple point), and a 
displacement to the left  for $h=[24,24,0]$. Level $i$ population densities  depend both on $\lambda$ and $h$, and suffer changes when approaching a critical point (see Figure \ref{fig5} later on).

\section{U(L) Coherent quasi-classical states and their operator expectation values}\label{cohsec}

$U(L)$ Coherent states $|h,U\ra$ turn out to be excellent variational states that reproduce the structure and mean energy of lowest-energy states inside each symmetry sector $h$ 
(see e.g \cite{octavio} for for the case of the $L=2$ level LMG model and \cite{PhysRevA.92.053843,PhysRevA.94.013802} for 
a system of $N$ indistinguishable atoms of $L$ levels interacting dipolarly with
$\ell$ modes of an electromagnetic field). 
They can be constructed by rotating each single particle state in, namely, the HW vector state $|\mbhw\ra$ by the same unitary matrix $U$. For example, for $L=3$ level atoms, an using the parametrization \eqref{GSU3} of a 
unitary matrix $U\in U(3)$, the  $|h,U\ra$ can be factorized as 
\be 
|h,U\ra=K_h(U)|h;\alpha,\beta,\gamma\},\label{cohu3}
\ee
where
\be
|h;\alpha,\beta,\gamma\}=e^{\beta S_{31}}e^{\alpha S_{21}} e^{\gamma S_{32}}|\mbhw\ra.\label{cohu3nonorm}
\ee
is the exponential action of lowering operators $S_{ij}, i>j$ on the HW state $|\mbhw\ra$ and 
\be
K_h(U)= |U|_1^{h_1-h_2} |U|_2^{h_2-h_3} |U|_3^{h_3}=\frac{u_1^{h_1}u_2^{h_2}u_3^{h_3}}{\ell_1^{(h_1-h_2)/2}\ell_2^{(h_2-h_3)/2}}\label{normcoh}
\ee
is a normalizing factor for $|h;\alpha,\beta,\gamma\}$ which depends on the lengths  \eqref{lengths} which appear  in  the products 
of first, second and third upper minors $|U|_i$ of $U=V\cdot\mathrm{diag}(u_1,u_2,u_3)$  in \eqref{GSU3}.  The  overlap 
\begin{align}
B_h(\bar\alpha',\bar\beta'&,\bar\gamma'; \alpha,\beta,\gamma)\equiv\{h;\alpha',\beta',\gamma'|h;\alpha,\beta,\gamma\} \nonumber\\  =&\left(1+\alpha  \bar{\alpha }'+\beta  \bar{\beta }'\right)^{h_1-h_2}  \nonumber \\ &\cdot
\left(1+\gamma  \bar{\gamma }'+(\beta -\alpha  \gamma ) (\bar{\beta }'-\bar{\alpha }' \bar{\gamma }')\right)^{h_2-h_3}\label{BergmanU3}
\end{align}
defines the so called reproducing Bergman kernel $B_h$.  Note that 
\be
B_h(\bar\alpha,\bar\beta,\bar\gamma; \alpha,\beta,\gamma)=\left|K_h(U)\right|^{-2}.\label{BK}
\ee
Coherent state expectation values $s_{ij}$ of the basic symmetry operators $S_{ij}$ can be easily computed through derivatives of the Bergman kernel as
\be
s_{ij}=\langle h, U|S_{ij}|h, U\rangle=\frac{\mathcal{S}_{ij}B_h(\bar\alpha,\bar\beta,\bar\gamma; \alpha,\beta,\gamma)}{B_h(\bar\alpha,\bar\beta,\bar\gamma; \alpha,\beta,\gamma) }.\label{sijsymb}
\ee
where $\mathcal{S}_{ij}$ is the  differential representation \eqref{difrelU3} of $S_{ij}$ on anti-holomorphic functions 
$\psi(\bar\alpha,\bar\beta,\bar\gamma)=\{h;\alpha,\beta,\gamma|\psi\ra$ (see Appendices  \ref{app1} and  \ref{app2} for a more detailed explanation). 
The explicit expression of the coherent state expectation values $s_{ij}$ can be seen in the equation \eqref{CSEV} of Appendix \ref{app1}. 
They will be very useful to compute the energy surface for each symmetry sector of the system in the next section. In Appendix \ref{app2} we study 
the interesting transformation properties of $U(3)$ coherent states under parity symmetry operations. These properties are strongly related to the degenerate 
structure of the ground state in the thermodynamic limit, as we shall see in the next section.

\section{Energy surface, phase diagram and spontaneously broken parity symmetry}\label{phasesec}

Let us consider a general $U(3)$ unirrep of shape $h=[h_1,h_2,h_3]$ given by the following proportions $\mu, \nu$
\bea &h_3=\nu N, \quad h_2=(1-\mu)(1-\nu)N, \quad h_1=\mu (1-\nu) N, & \nn\\
& \forall\, \nu\in[0,\frac{1}{3}], \quad \mu\in[\frac{1}{2},\frac{1-2\nu}{1-\nu}].& \eea
Note that the set of $U(3)$ unirreps labeled by $(\mu,\nu)$ is dense in the corresponding intervals as $N\to\infty$. 
The energy surface associated to a  Hamiltonian density $H$  inside the  Hilbert space sector $(\mu,\nu)$ 
is defined as the coherent state expectation value of the Hamiltonian density in the thermodynamic limit 
\begin{equation}
 E_{\mu,\nu}^U(\epsilon,\lambda)=\lim_{N\to\infty}\langle h, U|H|h, U\rangle.\label{energyU3}
\end{equation}
For the Hamiltonian density  \eqref{hamU3}, the energy surface becomes  
\begin{equation}
 E_{\mu,\nu}^U(\epsilon,\lambda)=\lim_{N\to\infty}\left(
 \frac{\epsilon(s_{33}-s_{11})}{N}-\frac{\lambda\sum_{i\not=j=1}^3 s_{ij}^2}{N(N-1)}\right),\label{energyU3-2}
\end{equation}
with $s_{ij}$ defined in \eqref{sijsymb} and calculated in \eqref{CSEV}.  Note that we have used the result \eqref{nofluct} which states that there are no fluctuations in the classical limit. 
This energy surface depends on the kind of unirrep $(\mu,\nu)$ (which become continuous parameters in the thermodynamic limit),  on the complex (phase space) coordinates  of $U$ (namely, $\alpha,\beta$ and $\gamma$) 
and on the control parameters $\epsilon$ and $\lambda$ related to the strength of interactions. Note that 
\be
E_{\mu,\nu}^U(\epsilon,\lambda)=\epsilon  E_{\mu,\nu}^U(1,\lambda/\epsilon),
\ee
which allows us to discuss the phase diagram in terms of the renormalized two-body  interaction strength $\tilde{\lambda}=\lambda/\epsilon$ for $\epsilon\not=0$. That is, we shall fix $\epsilon$ and measure energy and $\lambda$  in $\epsilon$ units.

Moreover, the fact that the unirreps $h=[h_1,h_2,h_3]$ and $h'=[h_1-h_3,h_2-h_3,0]$ 
are equivalent, under the point of view of $SU(3)$, introduces the following relation between energy surfaces
\be
E_{\mu,\nu}^U(\epsilon,\lambda)=(1-3\nu)E_{\tilde{\mu},0}^U(\epsilon,(1-3\nu)\lambda), \hspace{2mm}  \tilde{\mu}=\frac{\mu(1-\nu)-\nu}{1-3\nu},\label{enerequiv}
\ee
and therefore we can restrict ourselves to the analysis of the parent case $\nu=0, \mu\in[\frac{1}{2},1]$. The right end point $\mu=1$ 
corresponds to totally symmetric representations, associated to Young tableaux of a single row and four-dimensional phase spaces whose points are labeled by 
$\alpha, \beta\in \mathbb{C}$. Inserting the coherent state expectation values $s_{ij}$ of Eq.  \eqref{CSEV} into \eqref{energyU3-2} for the fully symmetric representation $[h_1,h_2,h_3]=[N,0,0]$, the 
corresponding energy surface turns out to be
\bea
E_{1,0}^{(\alpha,\beta)}(\epsilon,\lambda)&=&  \epsilon\frac{  \beta  \bar{\beta }-1}{
\alpha  \bar{\alpha }+\beta  \bar{\beta }+1} \label{enersym}\\ &&
-\lambda\frac{ \alpha ^2 \left(\bar{\beta }^2+1\right)+\left(\beta ^2+1\right) \bar{\alpha }^2+\bar{\beta }^2+\beta ^2
}{\left(\alpha  \bar{\alpha }+\beta  \bar{\beta }+1\right)^2}.\nonumber
\eea
Note that this energy surface is invariant under $\alpha\to-\alpha$, $\beta\to-\beta$, a symmetry  which is inherited from the discrete parity symmetry of the Hamiltonian 
\eqref{hamU3} already discussed at the end of Section \ref{LMGsec} and in Appendix \ref{app2}.

All representations with $h_1>h_2=h_3$ (that is, $\mu=\frac{1-2\nu}{1-\nu}$) can be reduced to this totally symmetric case $E_{1,0}^{(\alpha,\beta)}$; 
more precisely
\be
E_{\frac{1-2\nu}{1-\nu},\nu}^U(\epsilon,\lambda)=(1-3\nu)E_{1,0}^U(\epsilon,(1-3\nu)\lambda), \quad  0\leq \nu<1/3,
\ee
according to \eqref{enerequiv}. The left end point $\mu=1/2$ corresponds to the representations  
associated to rectangular Young tableaux of two equal  rows of $h_1=N/2=h_2$ boxes each and  four-dimensional  phase spaces whose  
points are labeled by two complex numbers $\gamma$ and $\beta'=\beta-\alpha\gamma$ [see the expression of $\ell_2$ in \eqref{lengths}]. 
The energy surface for this case is related to the totally symmetric case \eqref{enersym} through
\be
E_{\frac{1}{2},0}^U(\epsilon,\lambda)=\frac{1}{2} E_{1,0}^{(\gamma,\beta')}(\epsilon,\frac{\lambda}{2}),\quad \beta'=\beta-\alpha\gamma.
\ee
For intermediate  values 
$\mu\in(\frac{1}{2},1)$ the associated phase space is six-dimensional (a ``flag manifold'') and its points are labeled by three independent complex numbers $\alpha,\beta, \gamma$. 
The explicit expression of the energy surface for this case is  more bulky and we will not write  it down. 

Now we have to find the minimum energy 
\be E_\mu^{(0)}(\epsilon,\lambda)=\mathrm{min}_{U\in U(3)}E_{\mu,0}^U(\epsilon,\lambda)\label{Emin}\ee
for a parent representation $(\mu,0)$ with $\mu\in[\frac{1}{2},1]$. Eventually, we will find that the ground (minimal energy) state is always inside the 
totally symmetric sector $\mu=1$. 

The lowest-energy density $E_\mu^{(0)}(\epsilon,\lambda)$ inside each sector $\mu$ turns out to be 
\begin{widetext}
\begin{equation}
	E_\mu^{(0)}(\epsilon,\lambda)=
	\left\{
	\begin{matrix*}[l]
		\begin{Bmatrix*}[l]
			-\epsilon \mu   & &  & 0\leq \lambda \leq \frac{\epsilon }{2(1- \mu)} \\
			-\frac{1}{2} \left( \lambda  (1-\mu)^2+\frac{\epsilon ^2}{4\lambda }+(3 \mu-1)  \epsilon \right) && & \frac{\epsilon }{2(1- \mu)}\leq \lambda \leq \frac{3 \epsilon }{6 \mu -2} \\
			-\frac{2}{3} \lambda  (1-3 (1-\mu ) \mu)-\frac{\epsilon ^2}{2 \lambda }  & & & \lambda \geq \frac{3 \epsilon }{6 \mu -2}
		\end{Bmatrix*},
	&&&  \frac{1}{2}\leq \mu \leq \frac{2}{3}, \\ \\
	\begin{Bmatrix*}[l]
		-\epsilon \mu   & & & 0\leq \lambda \leq \frac{\epsilon }{4 \mu -2} \\
		2 \lambda  \mu (1-\mu) -\frac{(2 \lambda +\epsilon )^2}{8 \lambda }
		& && \frac{\epsilon }{4 \mu -2}\leq \lambda \leq \frac{3 \epsilon }{2} \\
		-\frac{2}{3} \lambda  (1-3 (1-\mu ) \mu)-\frac{\epsilon ^2}{2 \lambda }  && &  \lambda \geq \frac{3 \epsilon }{2}
	\end{Bmatrix*},
	&&&  \frac{2}{3}\leq \mu \leq 1.
	\end{matrix*}
	\right.
	\label{energyall}
\end{equation}
\end{widetext}

\begin{figure}[h]
\begin{center}
\includegraphics[width=\columnwidth]{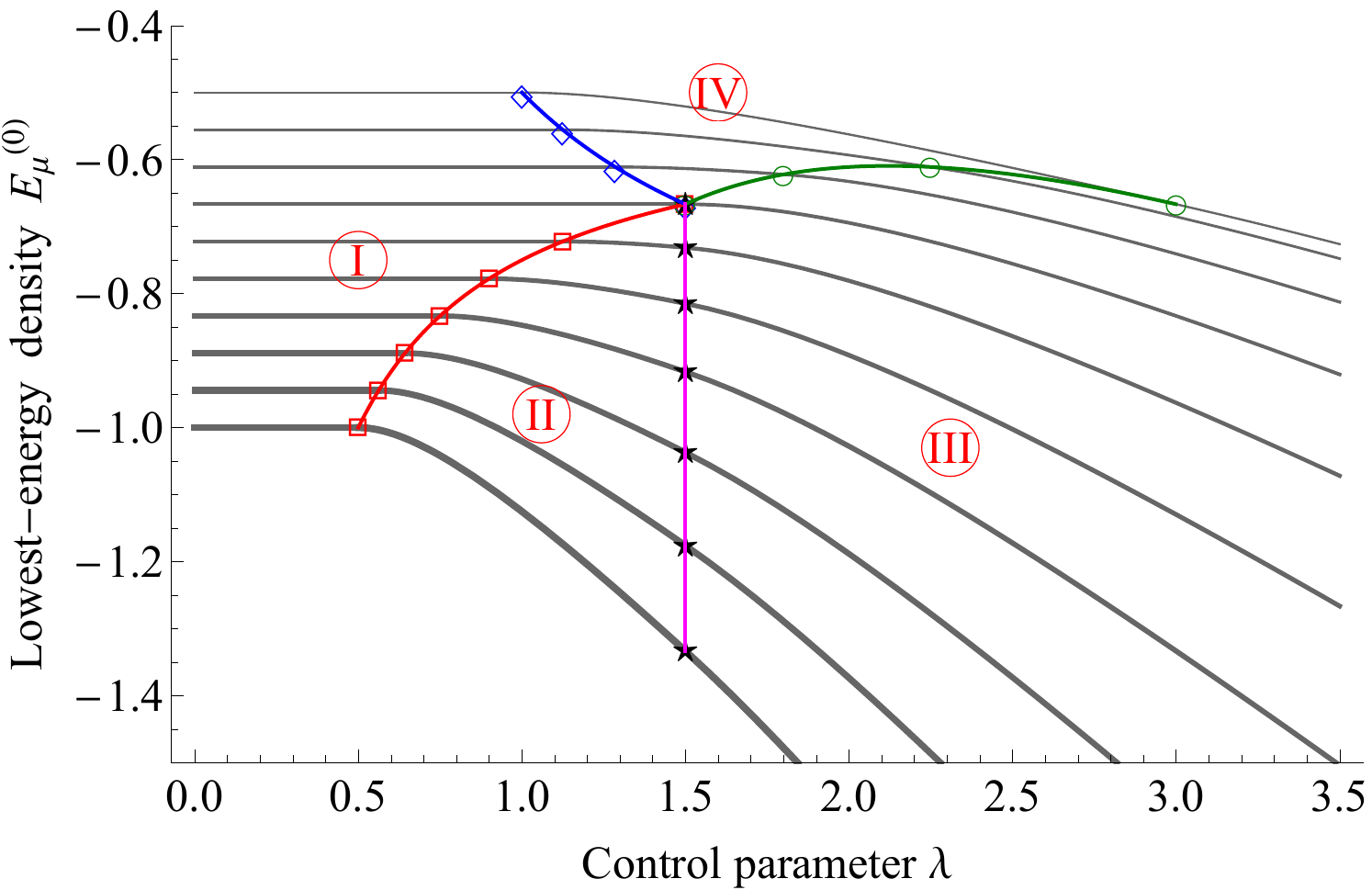}
\end{center}
\caption{Lowest-energy density $E_\mu^{(0)}(\epsilon,\lambda)$ as a function of the control parameter $\lambda$ (both in $\epsilon$ units) for ten different values of $\mu=n/18+8/18, n=1,\dots,10$, 
from $\mu=1/2$ (thinnest black curve) to $\mu=1$ (thickest black curve). Extending the control parameter space by $\mu$, the phase diagram exhibits four distinct quantum phases in the $\lambda$-$\mu$ plane that coexist at a 
quadruple point $(\lambda,\mu)_q=(3/2,2/3)$.  Curves of critical points separating two phases are depicted in color red,  blue, magenta  and green, according to formulas \eqref{lambdacrit}}
\label{fig4}
\end{figure}

\begin{figure}[h]
	\begin{center}
		\includegraphics[width=4cm]{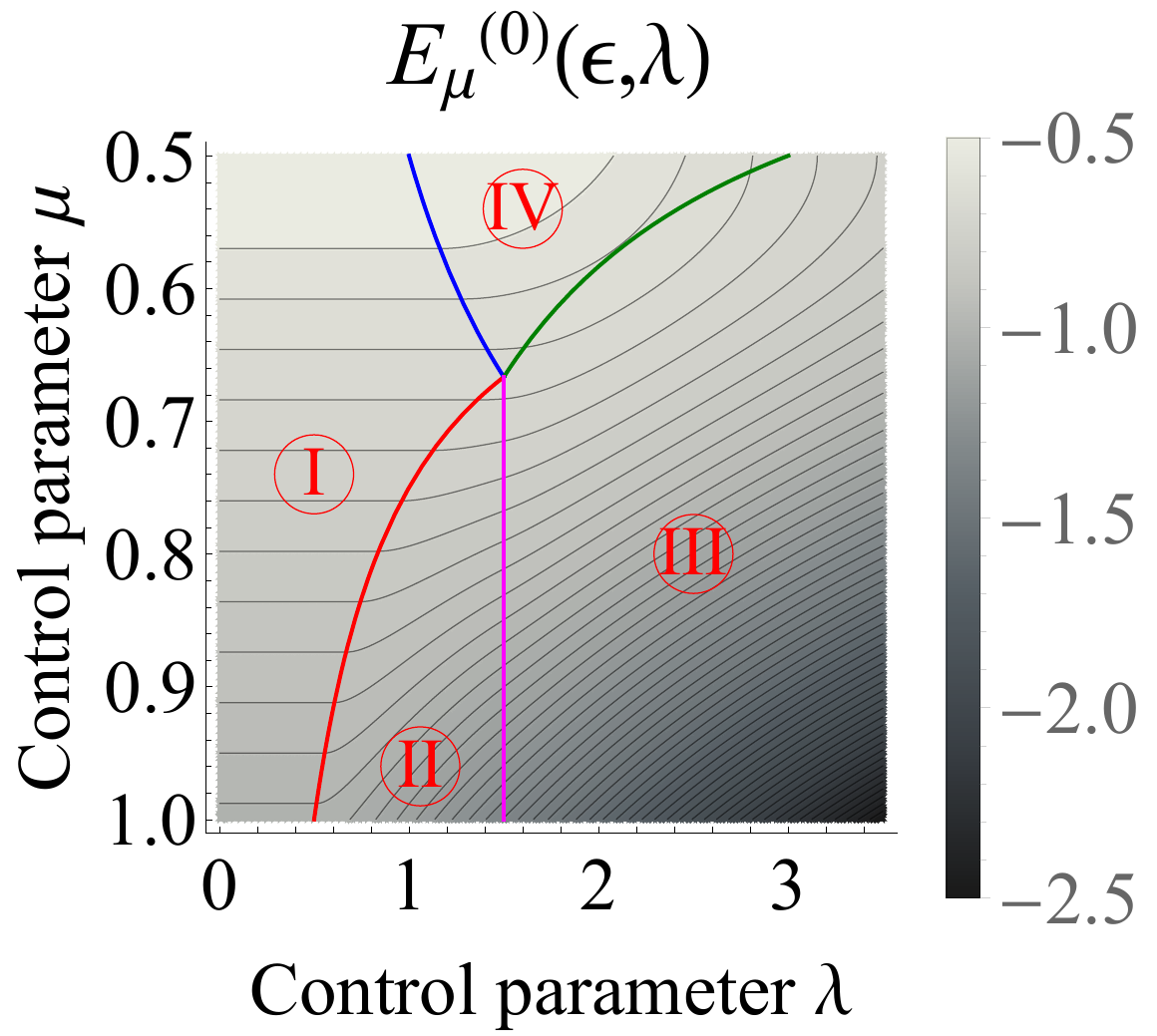}
		\includegraphics[width=4cm]{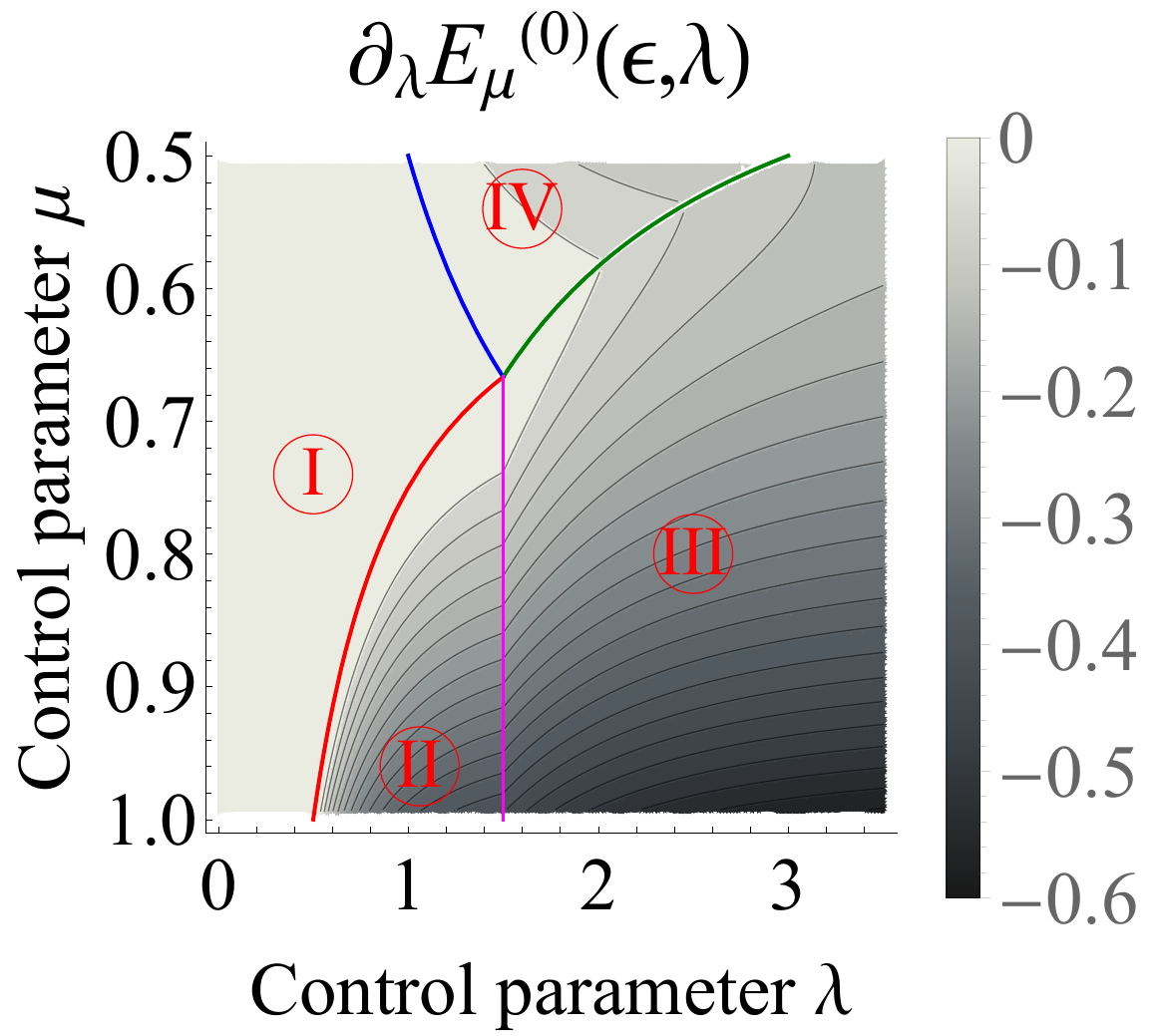}
		\includegraphics[width=4cm]{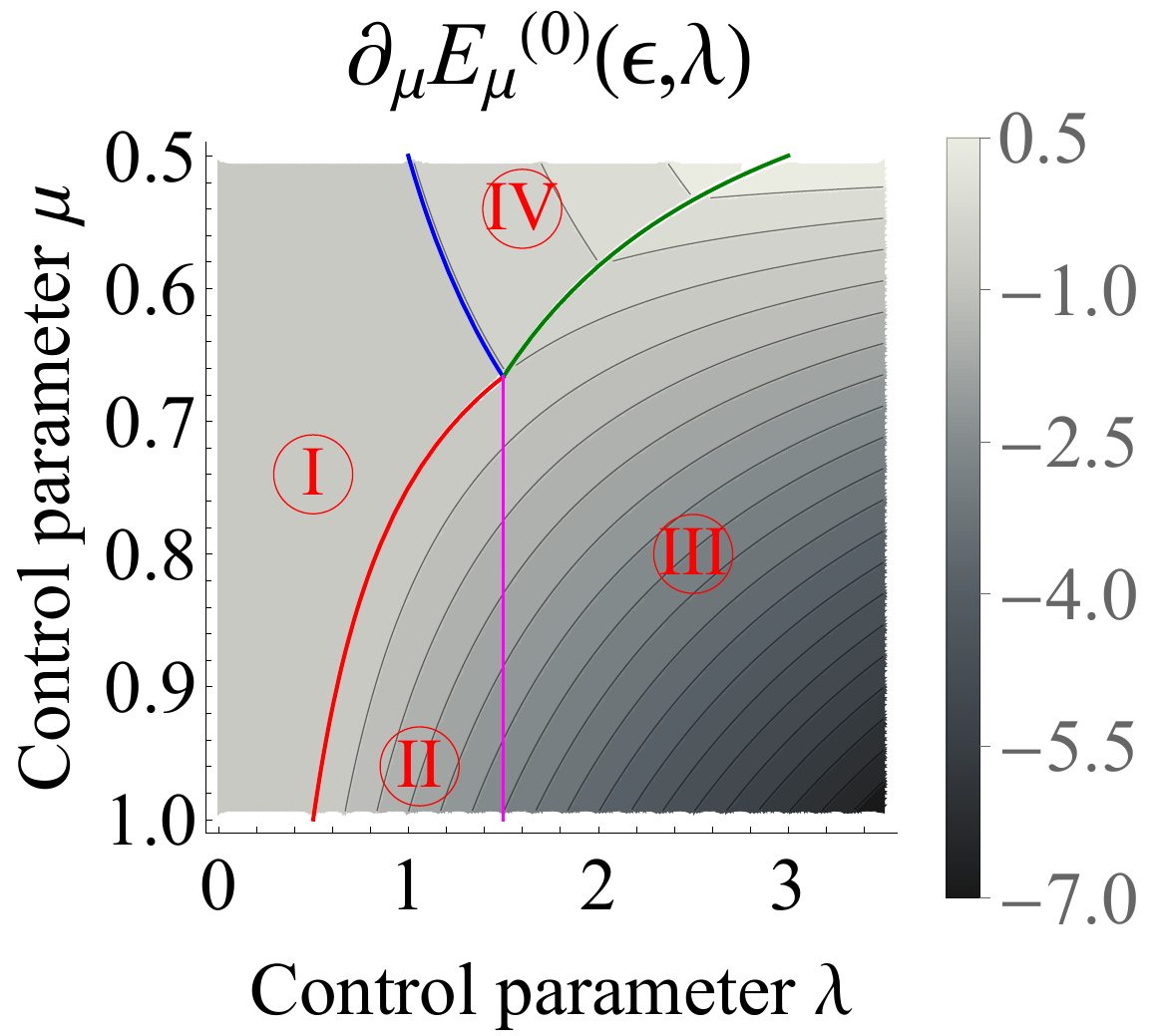}
		\includegraphics[width=4cm]{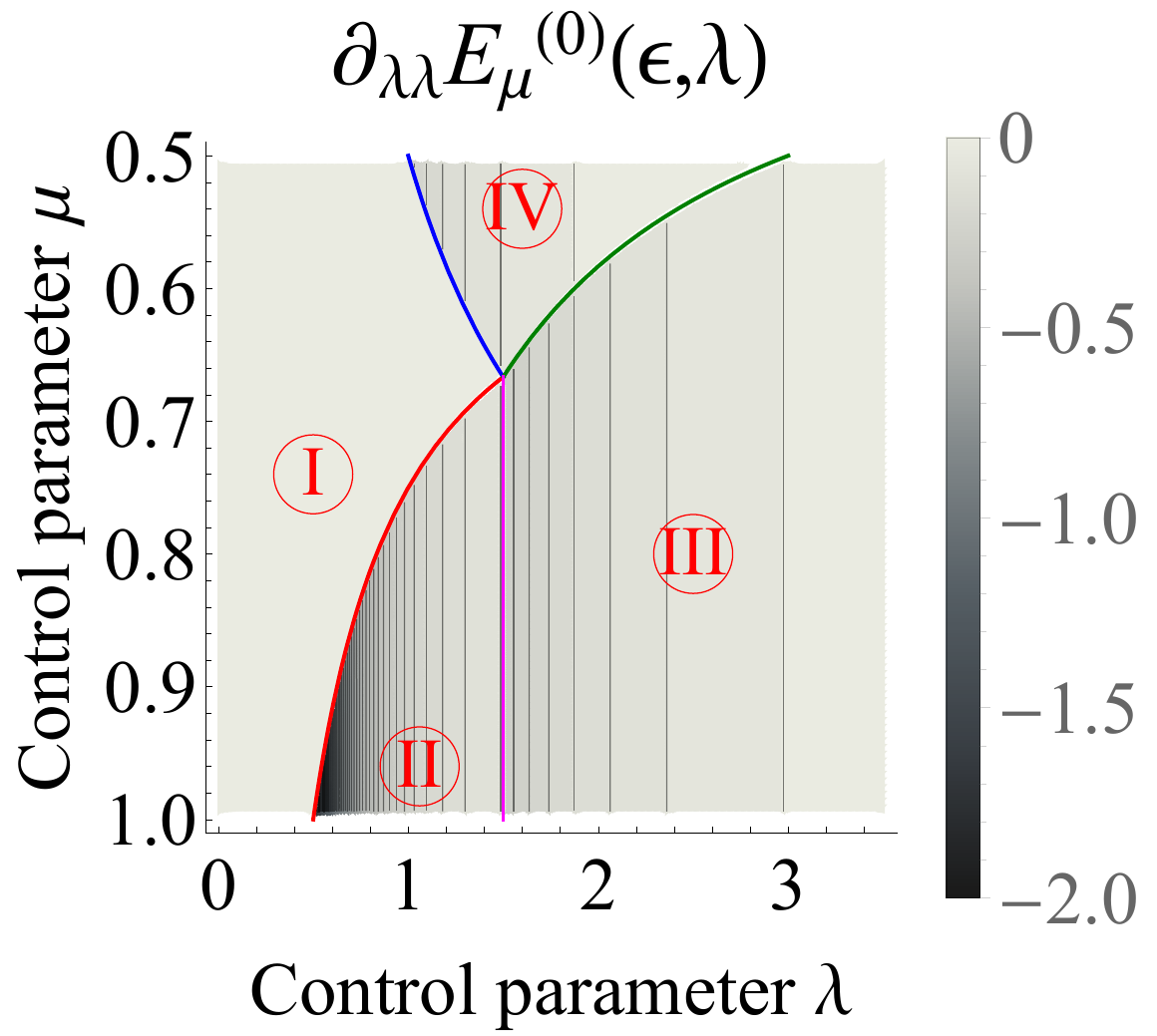}
		\includegraphics[width=3.9cm]{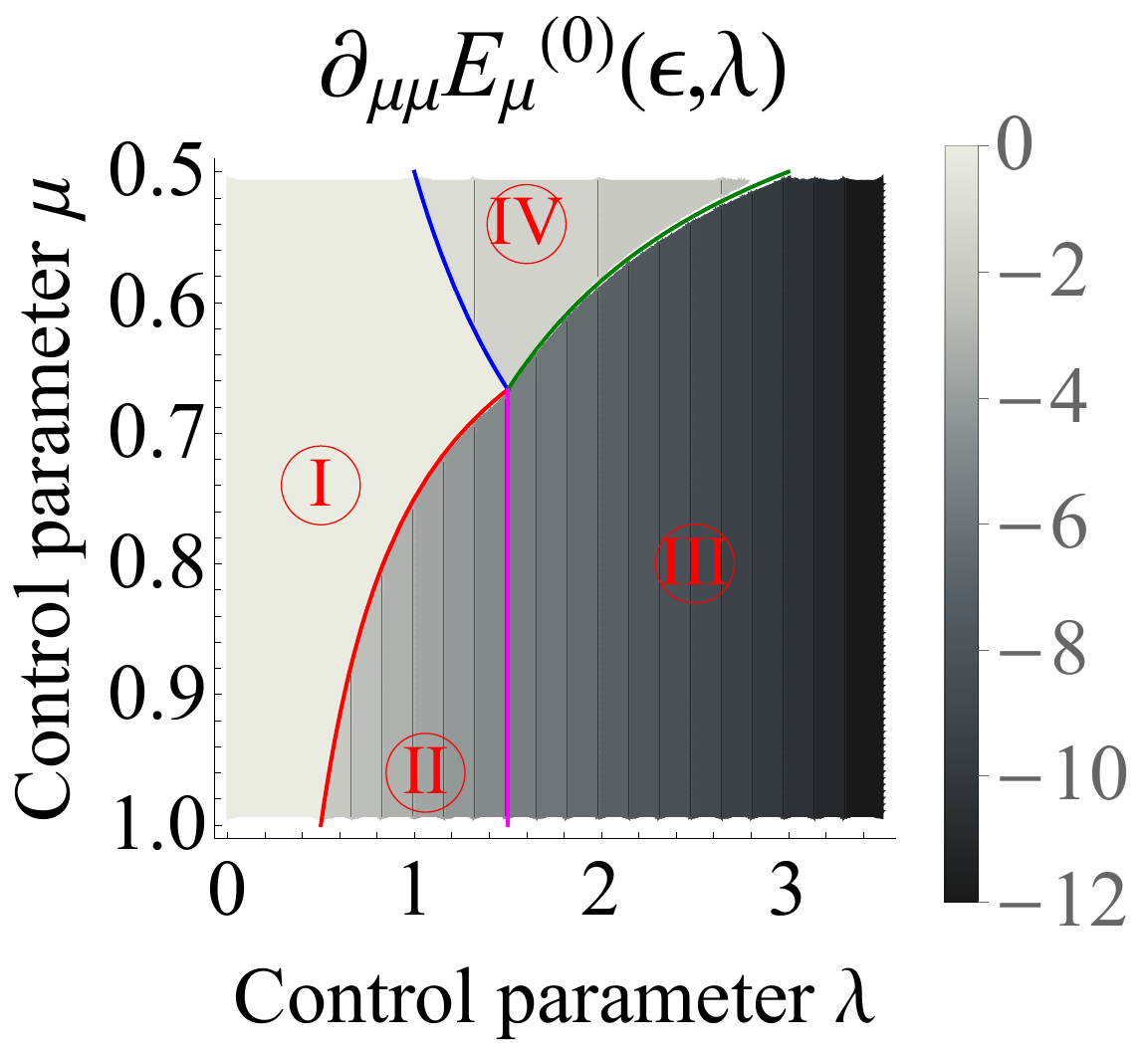}
		\includegraphics[width=4cm]{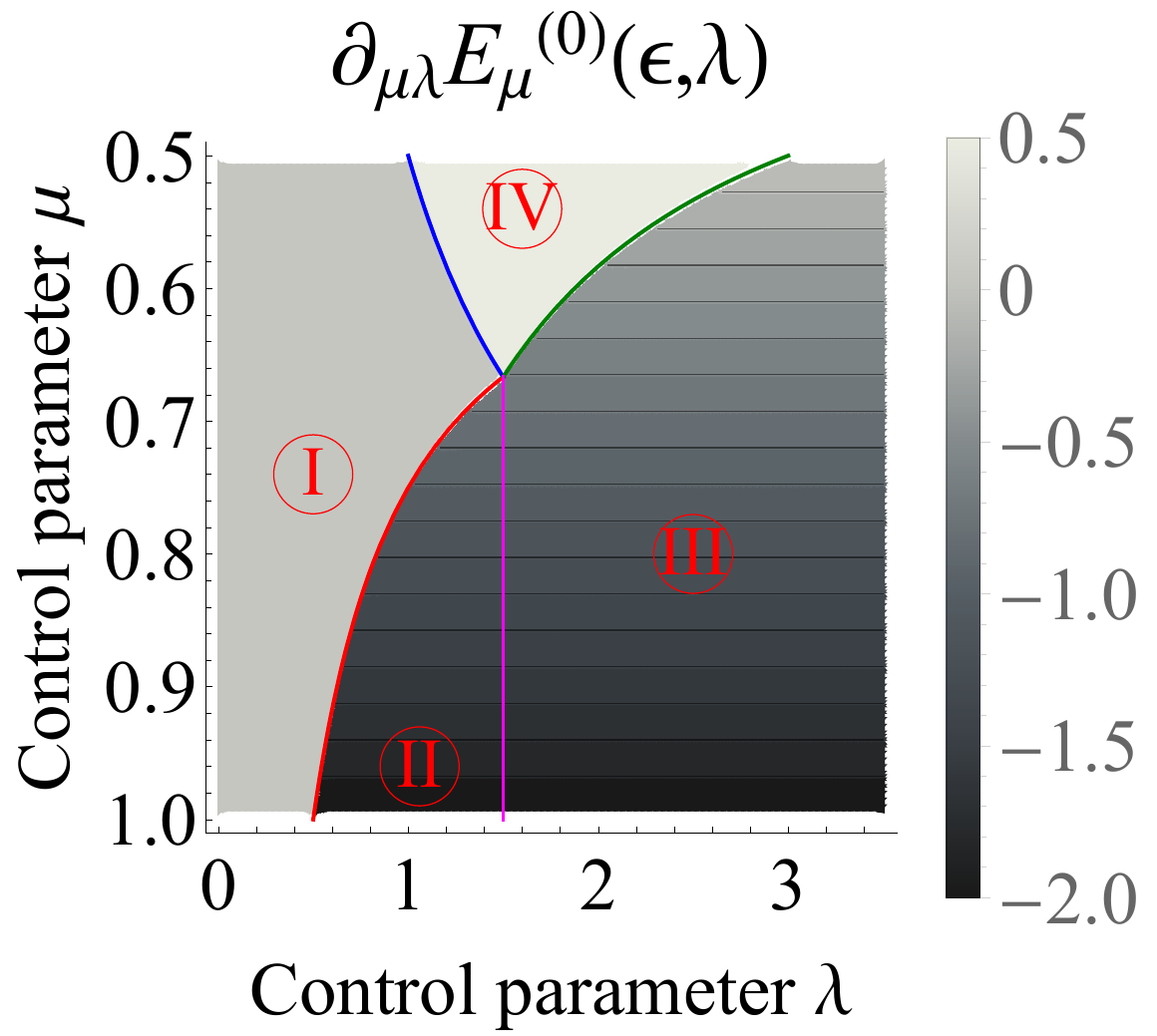}
	\end{center}
	\caption{Contour plots of $E_\mu^{(0)}(\epsilon,\lambda)$ and its first and  second derivatives in the phase diagram
	$(\lambda,\mu)$  (We use $\epsilon$ units for $E$ and $\lambda$). Critical curves (\ref{lambdacrit}) are shown, where the second derivatives are discontinuos.}
	\label{fig_phase}
\end{figure}

See Figure \ref{fig4} for a graphical representation of this energy as a function of $\lambda$ for several symmetry sectors $\mu$. Note that, in this new context of MSQPT, the sector parameter $\mu$ behaves as an additional 
control parameter. In fact, we can chose our initial quantum state inside a sector $\mu$ and Hamiltonian evolution will not take it out of this sector. We can also study quantum properties of arbitrarily close 
symmetry sectors $\mu$ and $\mu+\delta\mu$. Therefore, in this context, we extend the control parameter space $(\epsilon,\lambda)$ by $\mu$. Disregarding $\epsilon$, which only sets the scale (units), the 
phase diagram exhibits four distinct quantum phases (I, II, III and IV) in the $\lambda$-$\mu$ plane. These four quantum phases coexist at a 
quadruple point $(\lambda,\mu)_q=(3\epsilon/2,2/3)$, as it can be appreciated in Figures  \ref{fig4} and \ref{fig_phase}. We also represent curves of critical points separating two phases
\bea\begin{array}{rcllll}
\lambda^{(0)}_{\mathrm{I}\leftrightarrow\mathrm{II}}(\mu)&=&\frac{\epsilon}{4 \mu -2},&& \frac{2}{3}\leq \mu \leq 1, &(\mathrm{red})\\
\lambda^{(0)}_{\mathrm{II}\leftrightarrow\mathrm{III}}(\mu)&=&\frac{3 \epsilon}{2}, &&\frac{2}{3}\leq \mu \leq 1,&(\mathrm{magenta})\\
\lambda^{(0)}_{\mathrm{I}\leftrightarrow\mathrm{IV}}(\mu)&=&\frac{\epsilon}{2(1- \mu)},&& \frac{1}{2}\leq \mu \leq \frac{2}{3}, & (\mathrm{blue})\\
\lambda^{(0)}_{\mathrm{III}\leftrightarrow\mathrm{IV}}(\mu)&=&\frac{3 \epsilon}{6 \mu -2},&& \frac{1}{2}\leq \mu \leq \frac{2}{3}, & (\mathrm{green})
\end{array}\label{lambdacrit}
\eea
at which a second order QPT takes place in general. 

\begin{figure}[h]
	\begin{center}
		\includegraphics[width=7cm]{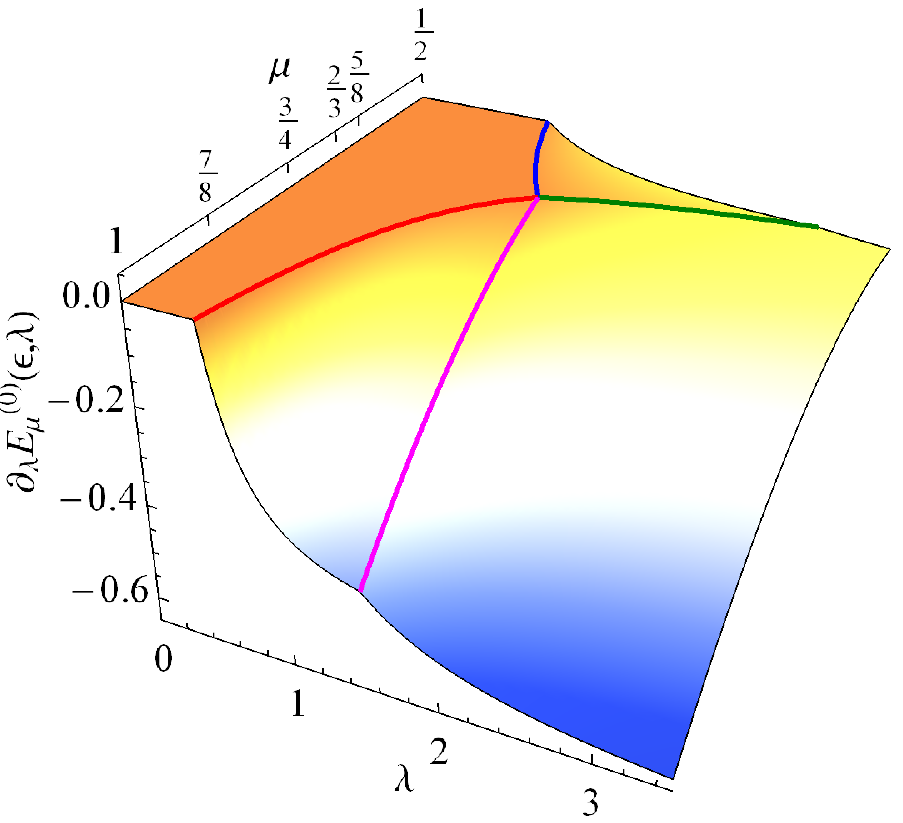}
		\includegraphics[width=7cm]{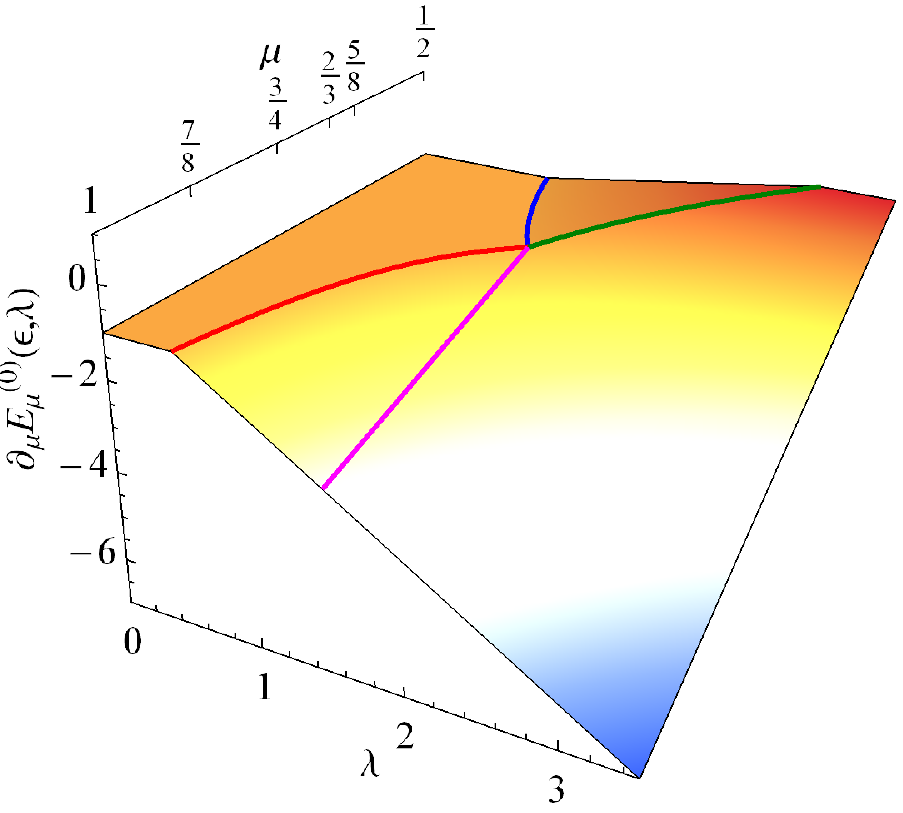}
	\end{center}
	\caption{3D plots of $\partial_{\lambda}E_\mu^{(0)}(\epsilon,\lambda)$ and $\partial_{\mu}E_\mu^{(0)}(\epsilon,\lambda)$  in the extended phase diagram
	$(\lambda,\mu)$  (We use $\epsilon$ units for $E$ and $\lambda$). 
	Critical curves (\ref{lambdacrit}) are shown, where a second order MSQPT occurs. 3D plots show that  both, $\partial_{\lambda}E_\mu^{(0)}$ and $\partial_{\mu}E_\mu^{(0)}$, are continuous. 
	The bottom plot shows that $\partial_{\mu}E_\mu^{(0)}$  attains its maximum at phase IV, where the density of $\mu$-levels with energy $E_\mu^{(0)}$  increases with 
	$\lambda$.}
	\label{3Dplots}
\end{figure}

To fully appreciate the nature of the phase-transitions, we show in Figure \ref{fig_phase} contour plots of the minimun energy 
$E_\mu^{(0)}(\epsilon,\lambda)$  and its first and second derivatives in the extended $(\lambda,\mu)$ phase diagram (i.e, considering both $\lambda$ and $\mu$ as control parameters). 
It is clear from the graphics that, while the first derivatives are continuous (see also Figure \ref{3Dplots}), the second derivatives are discontinuous at
the critical curves \eqref{lambdacrit} (at the curve  $\lambda^{(0)}_{\mathrm{II}\leftrightarrow\mathrm{III}}$  only 
$\partial_{\lambda\lambda}E_\mu^{(0)}$ is discontinuous). An interesting feature is the anomalous behavior  at phase $\mathrm{IV}$ where
$\partial_{\mu\lambda}E_\mu^{(0)}>0$, while it is non-positive in the rest of phases, this sign playing the role of 
an ``order parameter'' for the MSQPTs $\mathrm{I}\leftrightarrow\mathrm{IV}$ and $\mathrm{IV}\leftrightarrow\mathrm{III}$. This behavior can also be appreciated in Figure~\ref{fig4}, 
where one can see that energy curves of constant $\mu$ are parallel at region $\mathrm{I}$, move away from each other as $\lambda$ increases at regions $\mathrm{II}$ and $\mathrm{III}$, 
and get closer at region $\mathrm{IV}$. To better perceive it, we also represent in Figure \ref{3Dplots} 3D plots of $\partial_{\lambda}E_\mu^{(0)}(\epsilon,\lambda)$ and 
$\partial_{\mu}E_\mu^{(0)}(\epsilon,\lambda)$  in the extended phase diagram $(\lambda,\mu)$. As already said, we see that both first derivatives of the energy are continuous. Moreover, taking into account that the derivative $\partial_{\mu}E_\mu^{(0)}$ measures the density of $\mu$-levels with energy $E_\mu^{(0)}$, Figure \ref{3Dplots} (bottom panel) shows that the level density grows with $\lambda$ in phase IV, whereas it is non-increasing in the other phases, attaining its maximum value in phase IV. One can also 
perceive it in the fact that $\partial_{\mu\lambda}E_\mu^{(0)}>0$ in phase IV, as commented before. This discussion somehow connects with the traditional classification of ESQPTs characterized by a divergence in the density of excited states. We do not find any divergence of this kind (although we identify higher density level phases), but we must remind that our ``excited energy levels'' $E_\mu^{(0)}$ actually are the lower-energy levels inside each permutation symmetry sector $\mu$, the ground state corresponding to the symmetric sector $\mu=1$. This fact allows a variational analysis, both for QPTs and MSQPTs, in terms of coherent states $|h, U\rangle$ of $U(3)$.

\begin{figure}[h]
\begin{center}
\includegraphics[width=4cm]{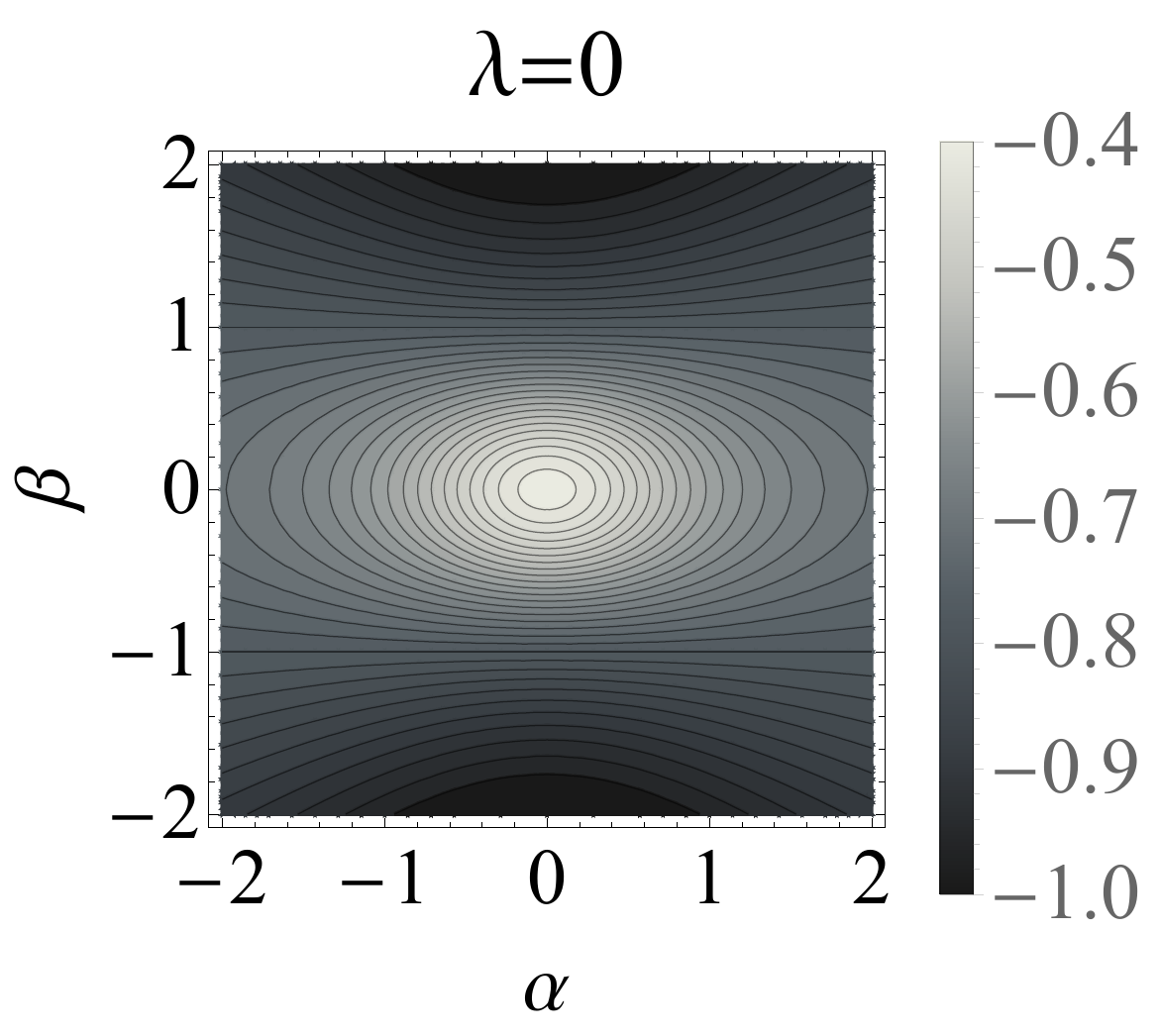}
\includegraphics[width=4cm]{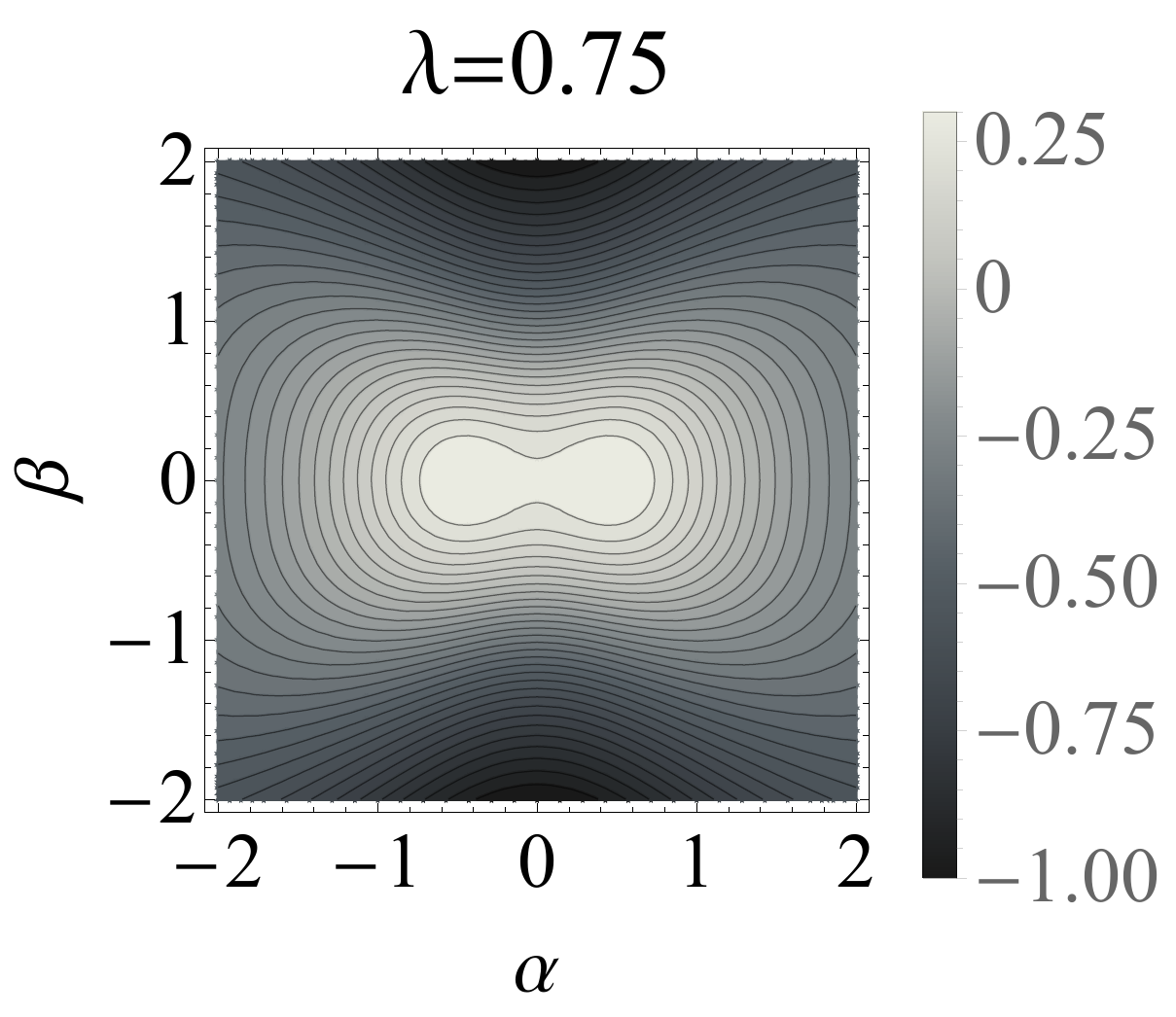}
\includegraphics[width=4cm]{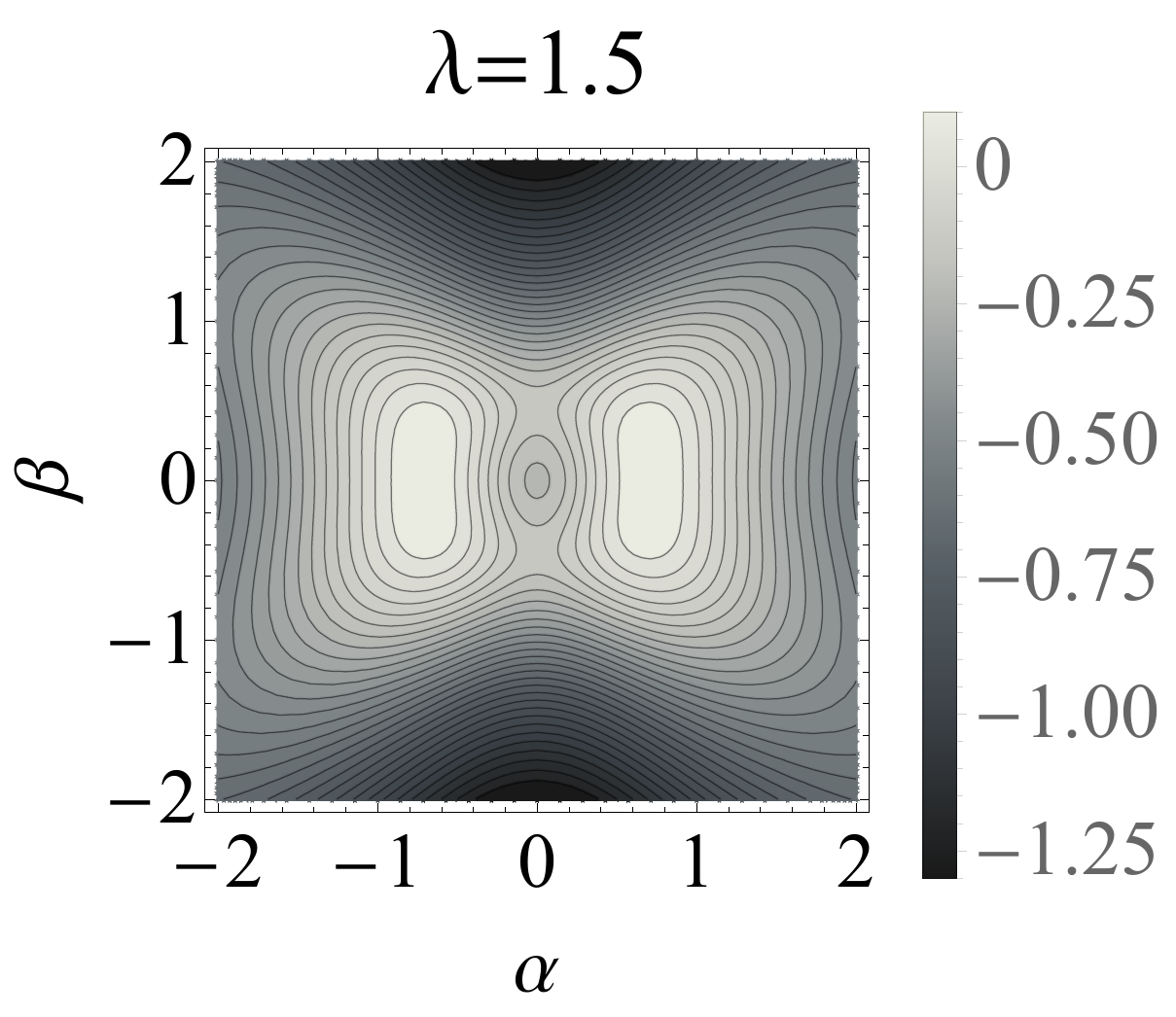}
\includegraphics[width=4cm]{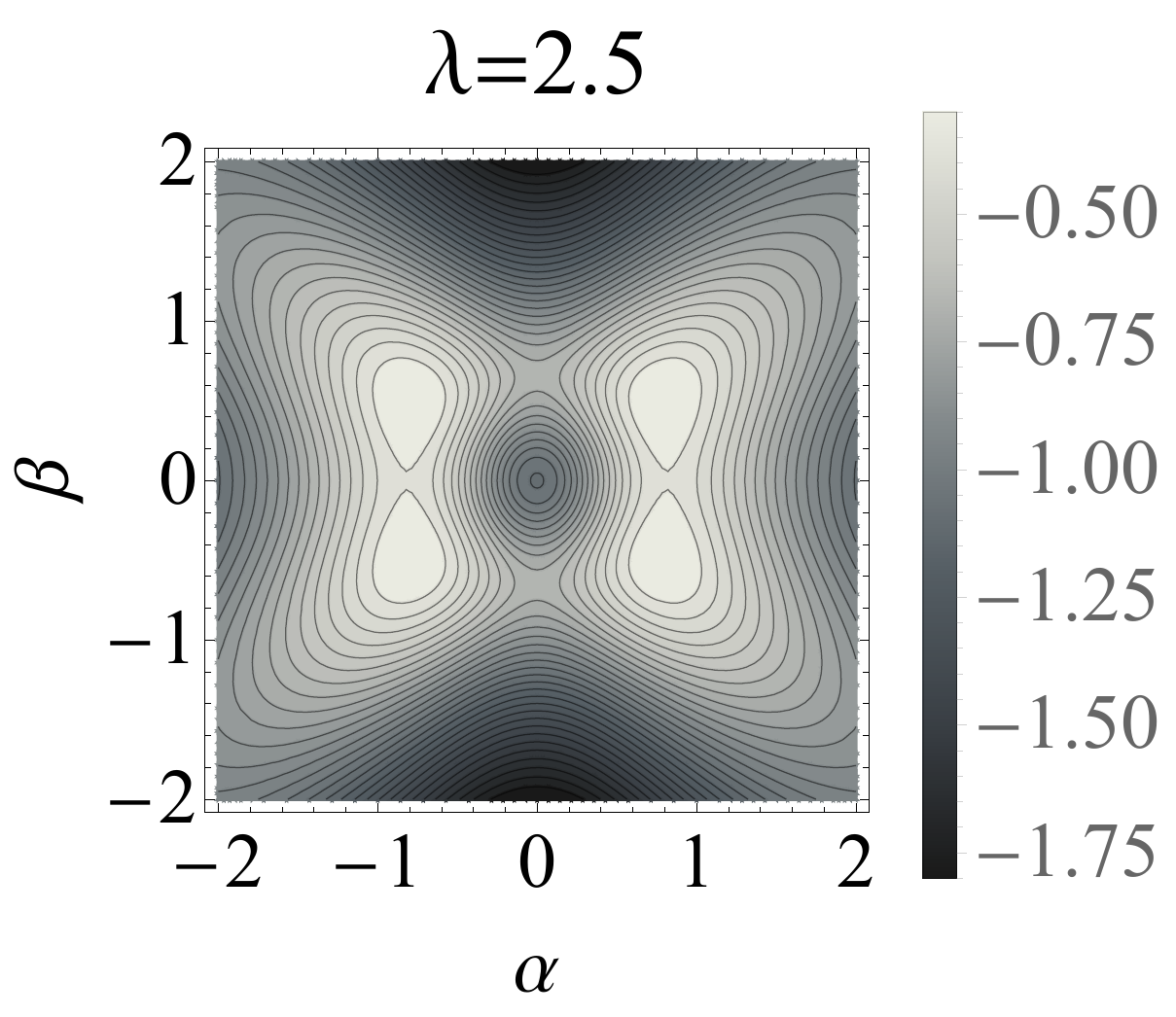}
\end{center}
\caption{Contour plot of the energy surface \eqref{enersym} of the fully symmetric case for real $\alpha$ and $\beta$, in the vicinity of the critical 
points $\lambda=\epsilon/2$ and $\lambda=3\epsilon/2$ (in $\epsilon$ units). Degenerate minima are perceived in light gray color. }
\label{fig3}
\end{figure}

Returning to our discussion on the minimization \eqref{Emin}, the lowest-energy state for general $\mu$ turns out to be highly degenerated. There are many 
phase space critical points $\alpha_0, \beta_0, \gamma_0$ with the same energy $E_\mu^{(0)}$ and the expressions are quite bulky. Therefore, we shall restrict ourselves from now on to the 
particular totally symmetric case $\mu=1$, which has a lower-dimensional phase space parameterized by $\alpha$ and $\beta$. 
The $\mu$-dependent lowest-energy \eqref{energyall}  simplifies for $\mu=1$ to 
\be
E_1^{(0)}(\epsilon,\lambda)=\left\{\begin{array}{lllr}
 -\epsilon,  && 0\leq \lambda \leq \frac{\epsilon }{2}, & \mathrm{(I)}\\
 -\frac{(2 \lambda +\epsilon )^2}{8 \lambda }, && \frac{\epsilon }{2}\leq \lambda \leq \frac{3 \epsilon }{2}, &  \mathrm{(II)} \\
  -\frac{4\lambda^2+3\epsilon ^2}{6 \lambda }, & &\lambda \geq \frac{3 \epsilon }{2}. &  \mathrm{(III)}\end{array}\right.\label{energysym}
\ee
Here we clearly distinguish the three different phases: I, II and III, and two second-order QPTs at $\lambda^{(0)}_{\mathrm{I}\leftrightarrow\mathrm{II}}=\epsilon/2$ and 
$\lambda^{(0)}_{\mathrm{II}\leftrightarrow\mathrm{III}}=3\epsilon/2$, respectively. 
The critical values of $\alpha$ and $\beta$ which make \eqref{enersym} minimum turn out to be real and their explicit expression is given by: 
\bea
\alpha_0^\pm(\epsilon,\lambda)&=&\pm\left\{\begin{array}{lll}
 0, && 0\leq \lambda \leq \frac{\epsilon }{2}, \\
 \sqrt{\frac{2\lambda- \epsilon }{2 \lambda +\epsilon }}, && \frac{\epsilon }{2}\leq \lambda \leq \frac{3 \epsilon }{2}, \\
 \sqrt{\frac{2\lambda }{2 \lambda +3 \epsilon }}, && \lambda \geq \frac{3 \epsilon }{2},
\end{array}\right.\nonumber\\
\beta_0^\pm(\epsilon,\lambda)&=&\pm\left\{\begin{array}{lll}
 0, & & 0\leq \lambda \leq  \frac{3 \epsilon }{2}, \\
 \sqrt{\frac{2 \lambda -3 \epsilon}{2 \lambda +3 \epsilon }}, & & \lambda \geq \frac{3 \epsilon }{2}. \end{array}\right. \label{critalphabeta}
\eea
Indeed, inserting \eqref{critalphabeta} into  \eqref{enersym} gives \eqref{energysym}. The location of these minima for the energy surface \eqref{enersym} can also be perceived by looking at the 
equipotential curves of Figure \ref{fig3}. Indeed, the real and imaginary parts of the complex phase-space variables $\alpha=x_\alpha+\ic p_\alpha$ and $\beta=x_\beta+\ic p_\beta$ can be seen as ``position'' $x$ and 
momenta $p$ (in dimensionless units). Minimum (potential) energy is attained for zero kinetic energy ($p$=0), i.e., real $\alpha$ and $\beta$. Looking at Figure \ref{fig3}, we perceive a single potential 
energy minimum in phase I, $0\leq \lambda/\epsilon\leq 1/2$, located at $\alpha=\beta=0$. In phase II, $1/2\leq \lambda/\epsilon\leq 3/2$, this single minimum degenerates into a double well potential. In phase III, 
$\lambda/\epsilon\geq 3/2$, we have a more degenerated case with a quadruple well potential, according to the critical values of $\alpha$ and $\beta$ in \eqref{critalphabeta}. 

As already commented,  this structure of degenerated minima is directly related with the spontaneous breakdown of the discrete parity symmetry of the Hamiltonian 
\eqref{hamU3} discussed at the end of Section \ref{LMGsec} and in Appendix \ref{app2}. Indeed, in the limit $N\to\infty$, the four coherent states $|\alpha_0^\pm,\beta_0^\pm\rangle$ attain the same minimum energy 
\eqref{energysym}. According to formula \eqref{paritycoh}, the parity operations $\hat\Pi_i=\Pi_ie^{-\ic\pi h_i}$ map between these four degenerate ground states; for example 
$\hat\Pi_1|\alpha_0^\pm,\beta_0^\pm\rangle=|\alpha_0^\mp,\beta_0^\mp\rangle$. Parity symmetry can still be restored by projecting any of the four $|\alpha_0^\pm,\beta_0^\pm\rangle$ degenerated 
ground states  onto the symmetric  (unnormalized) superposition 
\be
|\psi_0\}\equiv(1+\hat\Pi_1+\hat\Pi_2+\hat\Pi_3)|\alpha_0^\pm,\beta_0^\pm\rangle,
\ee
which remains invariant (even) under parity operations. These kind of ``parity-adapted coherent states'' have been extensively used in the literature and they are sometimes called ``Schr\"odinger cat states'', since 
they are a superposition of almost orthogonal semiclassical (coherent) states. The restoration of parity is convenient when one wants to compare between variational and (finite $N$) numerical calculations. For example, 
see \cite{PhysRevA.84.013819,L_pez_Pe_a_2015} for their use in the Dicke model of superradiance (two and three level atoms, respectively, interacting with one-mode radiation), 
\cite{Romera_2014,Calixto_2017} for the 2-level LMG model and \cite{Calixto_2012,PhysRevA.89.032126} for vibron models of molecules. We shall exploit this parity-symmetry restoration in future works.

\begin{figure}[h]
\begin{center}
\includegraphics[width=\columnwidth]{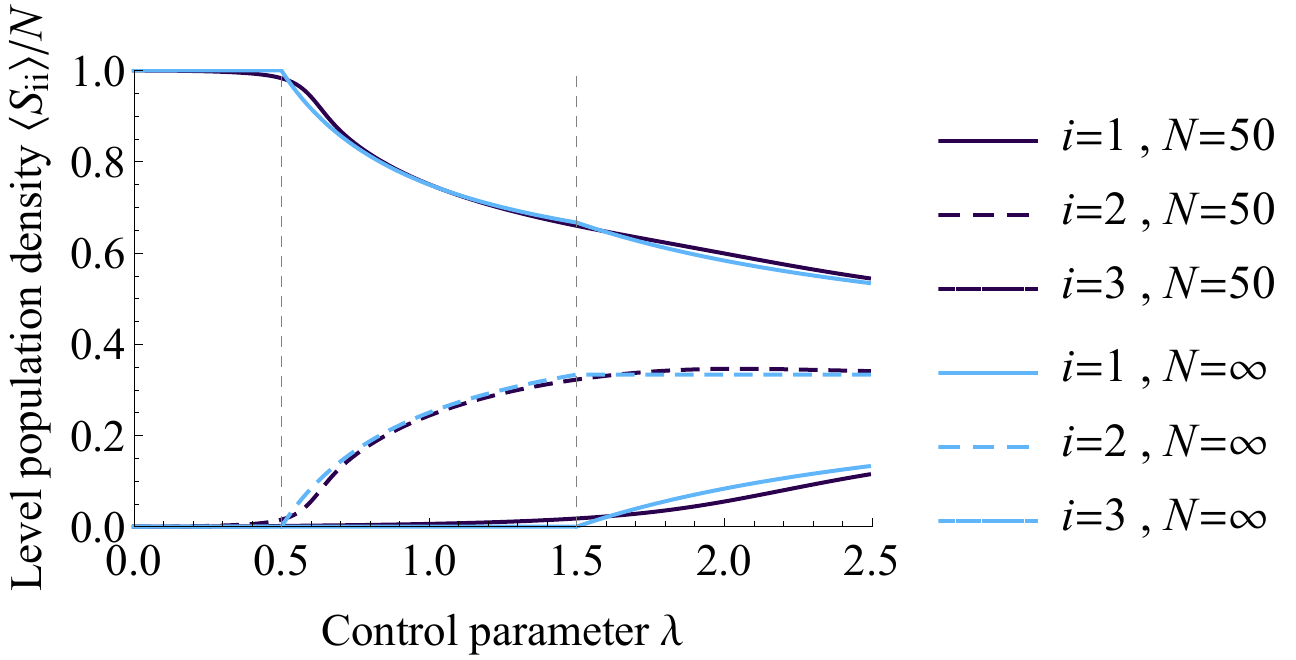}
\end{center}
\caption{Level population densities $p_{ii}^{(0)}, i=1,2,3,$ in eq. \eqref{levelpop},  corresponding to the fully-symmetric ground state in the thermodynamic 
limit as a function of the control parameter $\lambda$ (in $\epsilon$ units). Critical points indicate a change of behavior and are marked with vertical grid lines. 
We also compare with the finite case $N=50$, which already captures the critical behavior.}
\label{fig5}
\end{figure}

To finish, let us discuss other interesting order parameters of the QPT like the population density of each level. 
In Figure \ref{fig5} we show the level population density of the fully symmetric ground state ($h=[N,0,0]$) in the thermodynamic limit
\be
p_{ii}^{(0)}(\epsilon,\lambda)=\lim_{N\to\infty}\frac{\langle h,U_0|S_{ii}|h,U_0\rangle}{N}.
\ee
It can be explicitly calculated by using the expressions of the average values $s_{ii}$ of the operators $S_{ii}$ given in formulas \eqref{sijsymb} and \eqref{CSEV}, and then  evaluating them  
at the critical points \eqref{critalphabeta} as
\bea
p_{11}^{(0)}(\epsilon,\lambda)&=& \frac{1}{\ell_1(\alpha_0^\pm,\beta_0^\pm)}=\left\{\begin{array}{lll}
 1, && 0\leq \lambda \leq \frac{\epsilon }{2}, \\
 \frac{1}{2}+\frac{\epsilon}{4\lambda}, && \frac{\epsilon }{2}\leq \lambda \leq \frac{3 \epsilon }{2}, \\
 \frac{1}{3}+\frac{\epsilon}{2\lambda}, && \lambda \geq \frac{3 \epsilon }{2},
\end{array}\right.\nonumber\\
p_{22}^{(0)}(\epsilon,\lambda)&=&\frac{|\alpha_0^\pm|^2}{\ell_1(\alpha_0^\pm,\beta_0^\pm)}=\left\{\begin{array}{lll}
 0, && 0\leq \lambda \leq \frac{\epsilon }{2}, \\
 \frac{1}{2}-\frac{\epsilon}{4\lambda}, && \frac{\epsilon }{2}\leq \lambda \leq \frac{3 \epsilon }{2}, \\
 \frac{1}{3}, && \lambda \geq \frac{3 \epsilon }{2},
\end{array}\right.\nonumber\\
p_{33}^{(0)}(\epsilon,\lambda)&=& \frac{|\beta_0^\pm|^2 }{\ell_1(\alpha_0^\pm,\beta_0^\pm)}=\left\{\begin{array}{lll}
 0, && 0\leq \lambda \leq \frac{3\epsilon }{2}, \\
 \frac{1}{3}-\frac{\epsilon}{2\lambda}, && \lambda \geq \frac{3 \epsilon }{2}.
\end{array}\right.\nonumber\\ \label{levelpop}
\eea
Note that both energy-degenerate values $\pm$ also give the same population density, since $s_{ii}$  depend on squared modulus. 
We also compare in Figure \ref{fig5} with the population densities obtained for finite $N=50$, which already capture the critical behavior. We perceive a 
different population structure in each phase. In Phase I, $0\leq \lambda/\epsilon\leq 1/2$, all atoms are in level 1. In phase II, $1/2\leq \lambda/\epsilon\leq 3/2$, level 2 starts populating 
at the expense of level 1. Finally, level 3 begins to populate in phase III, $\lambda/\epsilon\geq 3/2$.

\section{Conclusions and outlook}

Quantum Phase Transitions in many-body systems usually presuppose the indistinguishability of the particles that compose the system, thus restricting the study to the 
fully symmetric representation ($\mu=1$ in our parametrization), but this should not be the more general situation. In this article we have analyzed the role played by 
other mixed symmetry sectors ($\mu\not=1$) in the thermodynamic limit $N\to\infty$ for a 3-level LMG model with $U(3)$ dynamical symmetry. We have seen that every lowest-energy 
state belonging to a given symmetry sector $\mu$ undergoes a QPT and the critical point $\lambda$ depends on $\mu$. This fact motivates the notion of 
{\em Mixed Symmetry Quantum Phase Transition} (MSQPT), leading to  an extended phase diagram in an enlarged control parameter space including $\mu$. Therefore, 
the system undergoes abrupt changes, not only for some critical values of the control parameters $\lambda$, but also for some critical values of the symmetry sector $\mu$. 
We also find that $\mu=2/3$ (the ``octet'' for $N=3$ particles/``quarks'') represents a quadruple point where four distinct phases coexist.  A numerical treatment for large (but finite) size $N$ 
gives some QPT precursors, like information-theoretic measures (fidelity-susceptibility) and level population densities, which anticipate some mean-field calculations 
using coherent (quasi-classical) states in the limit $N\to\infty$.

It would be interesting to further investigate the possible overlap between the proposed notion of MSQPT and the existing notion of Excited State Quantum Phase Transition (ESQPT) already present in the literature \cite{ESQPT,relano}, 
as we also find variability in the energy $\mu$-level density $\partial_{\mu}E_\mu^{(0)}$. 

Regarding the possibility  to exploit permutation symmetry for quantum technological prospects, we could mention for example some recent proposals commenting on thermodynamic 
advantages of bosonic over fermionic symmetry \cite{PhysRevE.101.012110}, or the role of  mixed symmetries in the quantum Gibbs paradox \cite{doi:10.1119/1.3584179,Yadin2020ExtractingWF}. 
The role of mixed symmetry in quantum  computation and  information theory also deserves our attention and will be investigated in future work.

Intermediate, fractionary, parastatistics also play a fundamental role in the quasiparticle zoo \cite{zoo}, that provides a deep understanding of complex phenomena in 
many-body and condensed matter physics. Recently, a proposal to describe composite fermions (in multicomponent fractional quantum Hall systems) in terms rectangular Young tableaux 
has been put forward \cite{CALIXTO1,multilayer} and showed to describe the quantum phases of bilayer quantum Hall systems with $U(4)$ dynamical symmetry \cite{PhysRevB.95.235302,Calixto_2018}. 
This is also an excellent area to explore the role of permutation symmetry.

\section*{Acknowledgments}
We thank the support of the Spanish MICINN  through the project PGC2018-097831-B-I00 and  Junta de Andaluc\'\i a through the projects SOMM17/6105/UGR, UHU-1262561 and FQM-381. 
JG also thanks MICINN for financial support from FIS2017-84440-C2-2-P. 
AM thanks the Spanish MIU for the FPU19/06376 predoctoral fellowship. We all thank Octavio Casta\~nos and Emilio Perez-Romero for their valuable collaboration in the early stages of this work.

\appendix 

\section{Differential realization of $S_{ij}$ and coherent state expectation values}\label{app1}

Let us justify the useful formula \eqref{sijsymb} and provide an explicit expression for the 
differential realization $\mathcal{S}_{ij}$ of the operators $S_{ij}$ on functions $\psi(\bar\alpha,\bar\beta,\bar\gamma)$. 
An alternative construction is also given in Appendix \ref{app2}. 

A group element $U'\in U(3)$ can be written as the exponential $U'=\exp(g'^{ij}S_{ij})$, with $g'^{ij}$ canonical coordinates at the identity. 
Using this, the coherent state expectation value \eqref{sijsymb} can also be written as
\be
\langle h, U|S_{ij}|h, U\rangle=\left.\frac{\partial \langle h, U|U'|h, U\rangle}{\partial g'^{ij}}\right|_{U'=1}.
\ee
Since $U'|h, U\rangle=|h, U'U\rangle$ and \[\langle h, U|h, U'U\rangle=\overline{K_h(U)} K_h(U'U) B_h(U^\dag;U'U),\] [with $K_h$ and $B_h$ in (\ref{normcoh},\ref{BergmanU3})], 
applying the chain rule of differentiation, the relation \eqref{BK} and the identification
\be
(g^{ij})=\left(
\begin{array}{ccc}
 u_1 & -\ac & -\bc \\
 \alpha  & u_2 & -\gc \\
 \beta  & \gamma  & u_3 \\
\end{array}
\right),
\ee
one finally arrives to the formula \eqref{sijsymb}. In order to apply the chain rule, one has to previously work out the group law $U''=U'U$, which means to write 
$g''^{ij}$ as a function of $g'^{ij}$ and $g^{ij}$. The corresponding group law is  quite cumbersome and we shall only write the final expression of the differential operators:
\bea
\mathcal{S}_{21}&=&\ac   (h_1-h_2)-(\bc -\ac  \gc ) \partial_{\gc}-\ac   \left(\bc  
\partial_{\bc}+\ac 
\partial_{\ac}\right), \nonumber\\ 
\mathcal{S}_{12}&=&\partial_{\ac},\nonumber\\ 
\mathcal{S}_{31}&=&(h_1-h_3)\bc + (h_3-h_2)\ac  \gc -
\gc  (\bc -\ac  \gc ) \partial_{\gc}\nonumber\\ 
&& -\bc  
\left(\bc  \partial_{\bc}+\ac   \partial_{\ac}\right),
\nonumber\\ 
\mathcal{S}_{13}&=&  \partial_{\bc},
\nonumber\\ 
\mathcal{S}_{32}&=& (h_2-h_3) \gc -
\gc^2 \partial_{\gc}+\bc  \partial_{\ac},
\nonumber\\ 
\mathcal{S}_{23}&=&  \partial_{\gc}+\ac   \partial_{\bc},
\nonumber\\ 
\mathcal{S}_{11}&=& h_1- \bc  \partial_{\bc}-\ac   
\partial_{\ac},
\nonumber\\ 
\mathcal{S}_{22}&=&h_2 +\ac   \partial_{\ac}-\gc  \partial_{\gc},
\nonumber\\ 
\mathcal{S}_{33}&=& h_3+\gc  \partial_{\gc}+\bc  \partial_{\bc}. \label{difrelU3}
\eea
With this, the corresponding expectation values \eqref{sijsymb} can be calculated and they are
\bea
s_{11}&=&  \frac{h_1}{\ell_1} +  \frac{h_2|\alpha +\beta  \gc |^2}{\ell_1\ell_2}+  \frac{ h_3|\beta -\alpha  \gamma|^2}{\ell_2}, \nonumber \\
s_{22}&=&  \frac{h_1|\alpha|^2}{\ell_1}   + \frac{ h_2|1-\alpha  \bc  \gamma +\beta  \bc|^2}{ \ell_1\ell_2}+  \frac{h_3 |\gamma|^2}{\ell_2},\nonumber\\
s_{33}&=&   \frac{ h_1 |\beta|^2  +  h_2(1+|\alpha|^2)}{\ell_1}+\frac{ h_3- h_2}{\ell_2},\nonumber\\
s_{12}&=&     \frac{(h_1-h_2) \alpha}{\ell_1}+ \frac{( h_2-h_3)\gc (\beta -\alpha  \gamma )}{\ell_2}, \nonumber \\
s_{13}&=&   \frac{( h_1- h_2)\beta  }{\ell_1}+\frac{( h_2- h_3) (\beta -\alpha  \gamma )}{\ell_2}, \nonumber \\
s_{23}&=&   \frac{( h_1- h_2)\ac  \beta  }{\ell_1}+\frac{( h_2- h_3)\gamma  }{\ell_2}
 \label{CSEV}
 \eea
and $s_{ij}=\bar{s}_{ji}$ for the reminder.

In the same way, the coherent state expectation value of operator higher powers can also be easily computed by repeated differentiation of the Bergman kernel. For example, 
for quadratic powers we have
\be
\langle h, U|S_{ij}S_{kl}|h, U\rangle=\frac{\mathcal{S}_{ij}\left(\mathcal{S}_{kl} B_h(\bar\alpha,\bar\beta,\bar\gamma; \alpha,\beta,\gamma)\right)}{B_h(\bar\alpha,\bar\beta,\bar\gamma; \alpha,\beta,\gamma) }.
\ee
However, to compute the energy surface \eqref{energyU3-2} we can restrict ourselves to expectation values \eqref{CSEV} since, in the thermodynamic/classical  limit $N\to\infty$,  quantum fluctuations disappear  and we have 
\begin{equation}
\lim_{N\to\infty}\frac{ \langle h, U|S_{ij}S_{kl}|h, U\rangle}{\langle h, U|S_{ij}|h, U\rangle\langle h, U|S_{kl}|h, U\rangle}= 1.\label{nofluct}
\end{equation}

\section{Parity symmetry operations on coherent states}\label{app2}
At the end of Section \ref{LMGsec}, we have seen that the parity operators $\Pi_i=\exp(\ic \pi S_{ii}), i=1,2,3$, are a symmetry of the 
Hamiltonian \eqref{hamU3}. This discrete symmetry is spontaneously broken in the thermodynamic limit, and degenerate ground  
states (``vacua'') arise in this limit. Coherent states \eqref{cohu3} are excellent variational states reproducing the 
ground state energy in the limit $N\to\infty$. Ground state degeneracy is perceived, for example, 
in the structure of multiple minima of the energy surface (\ref{energyU3-2},\ref{enersym}) depicted in Figure \ref{fig3} and calculated 
in \eqref{critalphabeta}. Note that critical values of the coherent state parameters $(\alpha,\beta,\gamma)$ appear in degenerate 
opposite pairs. Let us show that this is intimately related to the intrinsic parity symmetry of the Hamiltonian and discuss its 
consequences. We want to know the effect of a parity symmetry operation  $\Pi_i$ on a (non normalized) coherent state \eqref{cohu3nonorm}, 
that is
\be
\Pi_i|h;\alpha,\beta,\gamma\}=e^{\ic\pi S_{ii}}e^{\beta S_{31}}e^{\alpha S_{21}} e^{\gamma S_{32}}|\mbhw\ra.
\ee
All commutators $[S_{ii},S_{jk}]$, with $j>k$, are either zero or of the kind $[A,B]=\pm B$, for which we know that $[A,B^n]=\pm nB^n$ and 
$[A,e^{\alpha B}]=\pm \alpha Be^{\alpha B}$, which can be formally written as 
$[A,e^{\alpha B}]=\pm \alpha\partial_\alpha e^{\alpha B}$. In the same way, for the repeated commutator 
(adjoint action), we have
\[\mathrm{ad}_A^k(e^{\alpha B})\equiv [A,[A,\stackrel{k}{\dots}, [A,e^{\alpha B}]\dots]]=
\left(\pm \alpha\partial_\alpha\right)^ke^{\alpha B},\]
and therefore
\bea e^{\ic \theta A}e^{\alpha B}e^{-\ic \theta A}&=&\sum_{k=0}^\infty \frac{(\ic \theta)^k}{k!}\mathrm{ad}_A^k(e^{\alpha B})= 
e^{\pm \ic \theta\alpha\partial_\alpha }e^{\alpha B}\nonumber\\
&=&e^{e^{\pm \ic \theta}\alpha B}.
\eea
Taking into account the particular commutators $[S_{ii},S_{jk}], j>k$, setting $\theta=\pi$ and noting that 
$e^{\pm\ic \pi}=-1$ and $S_{ii}|\mbhw\ra=h_i|\mbhw\ra$, we finally arrive to
\bea
\Pi_1|h;\alpha,\beta,\gamma\}&=&e^{\ic\pi h_1}|h;-\alpha,-\beta,\gamma\},\nonumber\\
\Pi_2|h;\alpha,\beta,\gamma\}&=&e^{\ic\pi h_2}|h;-\alpha,\beta,-\gamma\},\nonumber\\
\Pi_3|h;\alpha,\beta,\gamma\}&=&e^{\ic\pi h_3}|h;\alpha,-\beta,-\gamma\}.\label{paritycoh}
\eea
From here we recover the fact that $\Pi_1\Pi_2\Pi_3=e^{\ic\pi N}$. We prefer to define the normalized parity operators $\hat\Pi_i=\Pi_ie^{-\ic\pi h_i}$, 
which verify $\hat\Pi_1\hat\Pi_2\hat\Pi_3=1$ and $\hat\Pi_i^{-1}=\hat\Pi_i$. Therefore, $\hat\Pi_3=\hat\Pi_1\hat\Pi_2$.

To finish, let us provide an alternative procedure to obtain the differential realization $\mathcal{S}_{ij}$ of $S_{ij}$ in \eqref{difrelU3}. 
Indeed, the property $[A,e^{\alpha B}]=\pm \alpha\partial_\alpha e^{\alpha B}$ implies that 
\bea
S_{11}|h;\alpha,\beta,\gamma\}&=&(h_1- \beta  \partial_{\beta}-\alpha   
\partial_{\alpha})|h;\alpha,\beta,\gamma\},\nonumber\\
S_{22}|h;\alpha,\beta,\gamma\}&=&(h_2 +\alpha   \partial_{\alpha}-\gamma  \partial_{\gamma})|h;\alpha,\beta,\gamma\},\nonumber\\
S_{33}|h;\alpha,\beta,\gamma\}&=&( h_3+\gamma  \partial_{\gamma}+\beta  \partial_{\beta})|h;\alpha,\beta,\gamma\},
\eea
which recovers the differential representation \eqref{difrelU3} of $S_{ii}$, for holomorphic functions this time. The deduction of non-diagonal 
$\mathcal{S}_{ij}, i\not=j$ follows a similar procedure, but it is a bit more involved and we shall not derive it here.

\bibliography{bibliografia.bib}
\bibliographystyle{apsrev4-1}

\end{document}